\documentclass[a4paper,twocolumn,11pt,unpublished]{quantumarticle}
\pdfoutput=1
\usepackage[utf8]{inputenc}
\usepackage[english]{babel}
\usepackage[T1]{fontenc}
\usepackage{amsmath}
\usepackage{hyperref}

\usepackage{tikz,lipsum,subfig,ulem,cancel}
\newcommand{\diff}{\mathrm{d}}

\newcommand{\bra}[1]{\left\langle{#1}\right|}
\newcommand{\ket}[1]{\left|{#1}\right\rangle}

\begin{document}

\title{Heisenberg-limited estimation robust to detector inefficiency in a multi-parameter Mach-Zehnder network with squeezed light}

\author{Dario Gatto}
\email{dario.gatto@port.ac.uk}
\affiliation{School of Mathematics and Physics, University of Portsmouth, Portsmouth PO1 3QL, United Kingdom}
\orcid{0000-0003-0252-2139}
\author{Paolo Facchi}
\orcid{0000-0001-9152-6515}
\affiliation{Dipartimento di Fisica and MECENAS, Universit\`a di Bari, I-70126  Bari, Italy}
\affiliation{INFN, Sezione di Bari, I-70126 Bari, Italy}
\author{Vincenzo Tamma}
\email{vincenzo.tamma@port.ac.uk}
\affiliation{School of Mathematics and Physics, University of Portsmouth, Portsmouth PO1 3QL, United Kingdom}
\affiliation{Institute of Cosmology and Gravitation, University of Portsmouth, Portsmouth PO1 3FX, United Kingdom}
\orcid{0000-0002-1963-3057}
\maketitle

\begin{abstract}
We propose a multi-parameter quantum metrological protocol based on a Mach-Zehnder interferometer with a squeezed vacuum input state and an anti-squeezing operation at one of its output channels. A simple and intuitive geometrical picture of the state evolution is provided by the marginal Wigner functions of the state at each interferometer output channel. The protocol allows to detect the value of the sum $\beta=\frac{1}{2}(\varphi_1+\varphi_2)+\theta_\mathrm{in}-\theta_\mathrm{out}$, of the relative phase $\theta_\mathrm{in}-\theta_\mathrm{out}$ between the two squeezers, and the average of the phase delays $\varphi_1,\varphi_2$ in the two arms of the interferometer. The detection sensitivity scales at the Heisenberg limit and, remarkably, is robust to 
the detector inefficiency.
\end{abstract}

\section{Introduction}
After Caves demonstrated, in a seminal work, that it is possible to reduce the quantum-mechanical noise of the signal in an interferometric experiment by fully harnessing the quantum nature of photons~\cite{caves81}, a great deal of interest has been invested in this endeavour, leading to the birth of the field of quantum metrology~\cite{yurke86,holland93,giovannetti04,giovannetti06,giovannetti11,pezze14,demkowicz15}.
In the near future, these technologies are expected to find applications in a wide range of settings. For instance, they could enhance the sensitivity in the mapping of inhomogenous magnetic fields~\cite{steinert10,hall12,pham11,seo07,baumgratz16}, phase imaging~\cite{humphreys13,liu16,yue14,knot16,gagatsos16,ciampini15}, quantum-enhanced nanoscale nuclear magnetic resonance imaging~\cite{eldredge18,arai15,lazariev15}, and long-distance clock synchronisation~\cite{komar14}.
Single-parameter quantum metrology, i.e.\ the problem of estimating a single parameter with quantum measurements has been extensively studied. By contrast the multi-parameter setting has remained vastly unexplored and has only recently become an attractive topic among the quantum physics research community~\cite{eldredge18,proctor17,boixo07,lang13,zhuang18,guo19,gramegna20,gramegna21}
because of its impact on the  development of quantum technologies.

In this work we describe an interferometric technique able to estimate with Heisenberg-limited sensitivity the combination $\beta=\frac{1}{2}(\varphi_1+\varphi_2)+\theta_\mathrm{in}-\theta_\mathrm{out}$ of the relative phase $\theta_\mathrm{in}-\theta_\mathrm{out}$ between two squeezers and the average of the upper phase $\varphi_1$ and the lower phase $\varphi_2$ in a Mach-Zehnder interferometer. We can thus estimate not only the average phase in the Mach-Zehnder if the relative phase of the squeezers is known, but also the relative phase of the squeezers, if on the other hand the average phase of the Mach-Zehnder is known.  The interferometer is also able to estimate the relative phase $\frac{1}{2}(\varphi_1-\varphi_2)$ between the two arms of the optical interferometer with a sensitivity scaling at the standard quantum limit. 
Remarkably, we demonstrate that the in both cases the effect of inefficient detectors, which might limit in practice the sensitivity of the scheme, only hinders the sensitivity by a constant factor, i.e.\ that our protocol is robust to external photon losses.

\begin{figure}[ht]
\centering
\includegraphics[width=.48\textwidth]{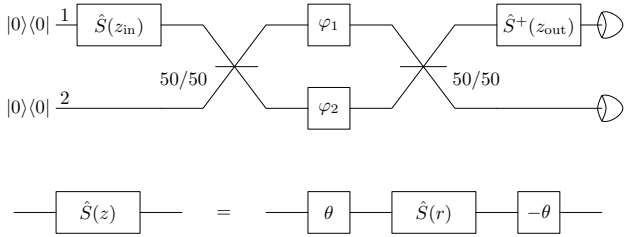}
\caption{Interferometric setup for the Heisenberg-limited  estimation of the combination $\beta=\frac{1}{2}(\varphi_1+\varphi_2)+\theta_\mathrm{in}-\theta_\mathrm{out}$. Squeezing operations are characterised by the complex squeezing parameter $z=r\,e^{i\theta}$. Specifically, a balanced Mach-Zehnder is preceded by a squeezer with squeezing parameter $z=z_{\mathrm{in}}=r\,e^{i\theta_\mathrm{in}}$, and followed by an anti-squeezing operation with squeezing parameter  $z=z_{\mathrm{out}}=r\,e^{i\theta_\mathrm{out}}$. On-off photodetectors are placed at the end of the interferometer.}
\label{fig:schematic}
\end{figure}

\section{State evolution through the interferometer}
Let us consider a balanced Mach-Zehnder interferometer (see the schematic in Fig.~\ref{fig:schematic}) where one of the input channels is fed with a squeezed vacuum state, characterised by the squeezing parameter $z_{\mathrm{in}}=r\,e^{i\theta_{\mathrm{in}}}$, and the other channel is left in the vacuum state.
Throughout the interferometer the system is in a Gaussian state described by the Wigner function~\cite{schleich}
\begin{equation}
 W_{\sigma}(\boldsymbol{\xi}) = \frac{e^{-\frac{1}{2}\boldsymbol{\xi}^T \,\sigma^{-1}\, \boldsymbol{\xi}}}{(2\pi)^2\sqrt{\det(\sigma)}},
 \label{eq:wigner_gaussian}
\end{equation}
where $\boldsymbol{\xi}=(x_1,p_1,x_2,p_2)^T$ is the 2-mode phase space variable, and $\sigma$ is the covariance matrix. 

At the input of the interferometer, the covariance matrix takes the form 
\begin{equation}
 \sigma_{\mathrm{in}} = \frac{1}{2}
\begin{pmatrix}
S(z_{\mathrm{in}})^2 & 0 \\
 0 & \openone_2
\end{pmatrix} ,
 \label{eq:squeezing_in}
\end{equation}
where
 \begin{equation}
 S(z) = S(r  e^{i\theta}) = e^{i \theta \sigma_y} \operatorname{diag}(e^{r}, e^{-r}) \, e^{-i \theta \sigma_y}
 \label{eq:squeezer}
 \end{equation}
is the matrix associated with the (one-mode) squeezing operation in phase space acting on the first channel, and $\sigma_y$ is the second Pauli matrix.
The mean photon number associated with the input state is $\frac{1}{2} \mathrm{tr} (\sigma_{\mathrm{in}}) -1 =
\sinh(r)^2 = N $. 

The action of the Mach-Zehnder interferometer on the quantum state is described by the covariance matrix transformation
\begin{equation}
 \sigma_{\mathrm{MZ}} = O_{\mathrm{MZ}} \sigma_{\mathrm{in}} O_{\mathrm{MZ}}^T ,
 \label{eq:interferometer_cov_transf}
\end{equation}
where the orthogonal and symplectic matrix $O_{\mathrm{MZ}}$ reads~\cite{gatto20} (see Appendix~\ref{sec:append_0})
\begin{align}
 O_{\mathrm{MZ}} = \begin{pmatrix}
 c_- \, e^{-i\varphi_+\sigma_y} &
  s_- \, e^{-i\left(\varphi_++\frac{\pi}{2}\right)\sigma_y}\\
 s_- \, e^{-i\left(\varphi_++\frac{\pi}{2}\right)\sigma_y} &
c_- \, e^{-i\varphi_+\sigma_y}
 \end{pmatrix},
 \label{eq:OMZ}
\end{align}
with $c_- = \cos(\varphi_-)$, $s_- = \sin(\varphi_-)$, and 
\begin{equation}
\varphi_\pm = (\varphi_1\pm\varphi_2)/2.
\end{equation}

The measurement operation at the output of the Mach-Zehnder interferometer is defined by an anti-squeezing operation on the first channel, characterised by the squeezing parameter $z_{\mathrm{out}}=re^{i\theta_{\mathrm{out}}}$, 
and by the consequent projection over the vacuum state via on-off photodetectors placed at the two output channels.
The operator associated with such measurement is the projector 
\begin{equation}
\label{eq:projector}
\hat{\Pi}=\hat{S}_1(z_\mathrm{out})|00\rangle\langle00|\hat{S}_1^\dag(z_\mathrm{out}),
\end{equation} 
where $\hat{S}_1(z) = e^{\frac{1}{2}(z\hat{a}_1^{\dag2}-z^*\hat{a}_1^2)}$ is the squeezing operator, and is associated with the Wigner function $W_{\sigma_\mathrm{out}}(\boldsymbol{\xi})$, with
\begin{equation}
 \sigma_\mathrm{out} 
 = \frac{1}{2}
 \begin{pmatrix}
S(z_{\mathrm{out}})^2 & 0 \\
 0 & \openone_2
\end{pmatrix} ,
 \label{eq:cov_meas}
\end{equation}
and $S(z_{\mathrm{out}})$ given by Eq.~\eqref{eq:squeezer}. The probability for ideal detectors to click is $1-P$, where
\begin{align}
P = \langle\hat{\Pi}\rangle &= (2\pi)^2 \int W_{\sigma_\mathrm{out}}(\boldsymbol\xi) \, W_{\sigma_\mathrm{MZ}}(\boldsymbol\xi) \,\diff^4\xi \nonumber\\
& = \det(\sigma_\mathrm{MZ}+\sigma_\mathrm{out})^{-1/2},
\label{eq:gaussian_overlap}
\end{align}
as can be seen by taking a simple Gaussian integral. The expectation value is taken on the Mach-Zehnder output state associated with the covariance matrix $\sigma_\mathrm{MZ}$ in Eq.~\eqref{eq:interferometer_cov_transf}.

We can track the evolution of the state throughout the interferometer by using the marginals of the correspondent Wigner function,
\begin{equation}
 W_{i}(x_i,p_i;\sigma) = \int W_{\sigma}(\boldsymbol{\xi})\,\diff x_j\diff p_j,
 \label{eq:reduced_wigners}
\end{equation}
at each channel $i,j=1,2$ and $i\neq j$, where $W_{\sigma}$ is given in Eq.~\eqref{eq:wigner_gaussian}. 
Of course, since the state in the two channels gets entangled by the interactions with the beam splitters, the information provided by the marginals cannot be expected to be complete. Nonetheless, as we shall see, the marginals are sufficient to provide a clear and intuitive physical picture. 

\begin{figure}
\subfloat[Channel 1, $\varphi_-=0$.]{\includegraphics[width=.5\columnwidth]{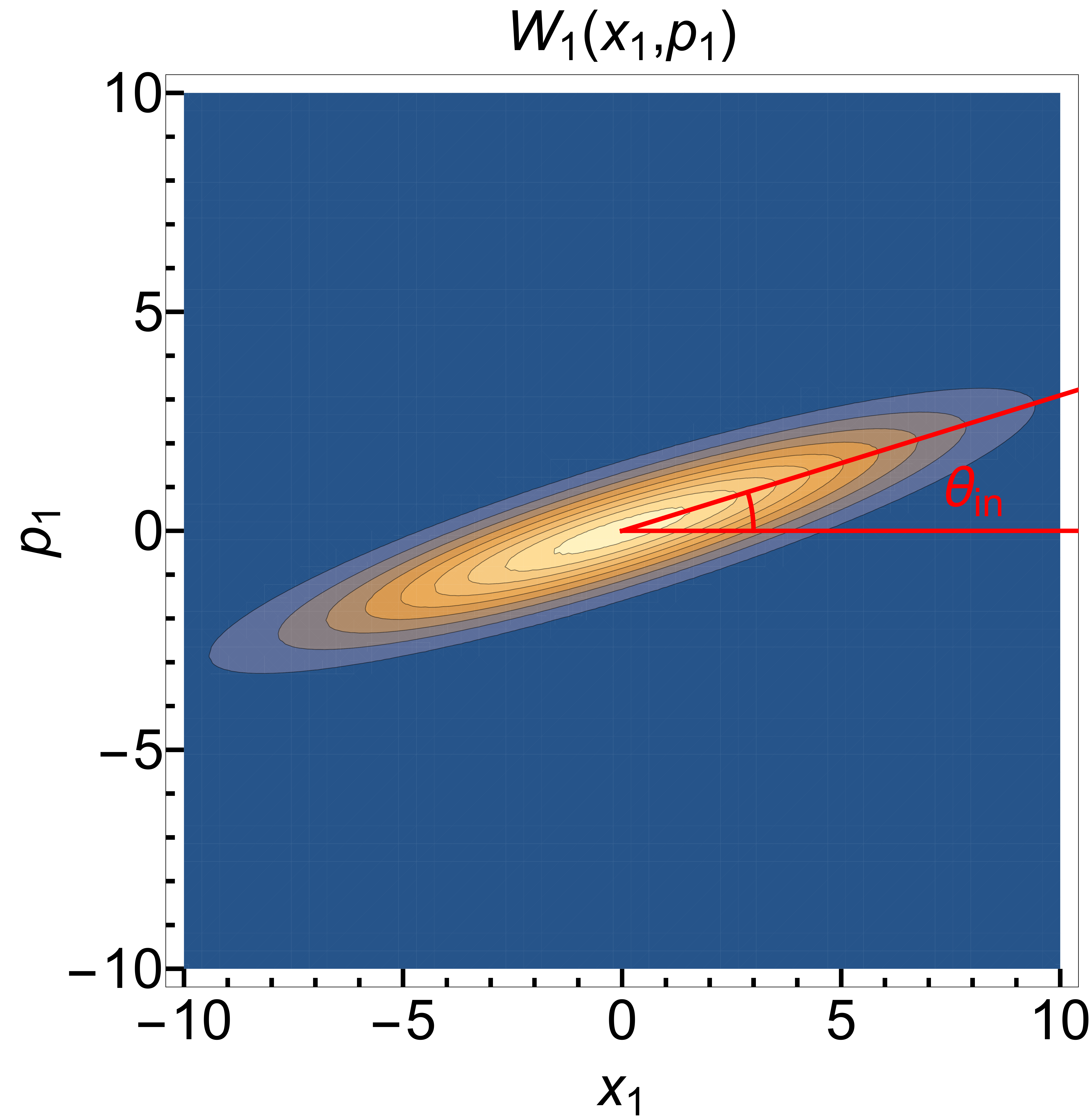}}
\subfloat[Channel 2, $\varphi_-=0$.]{\includegraphics[width=.5\columnwidth]{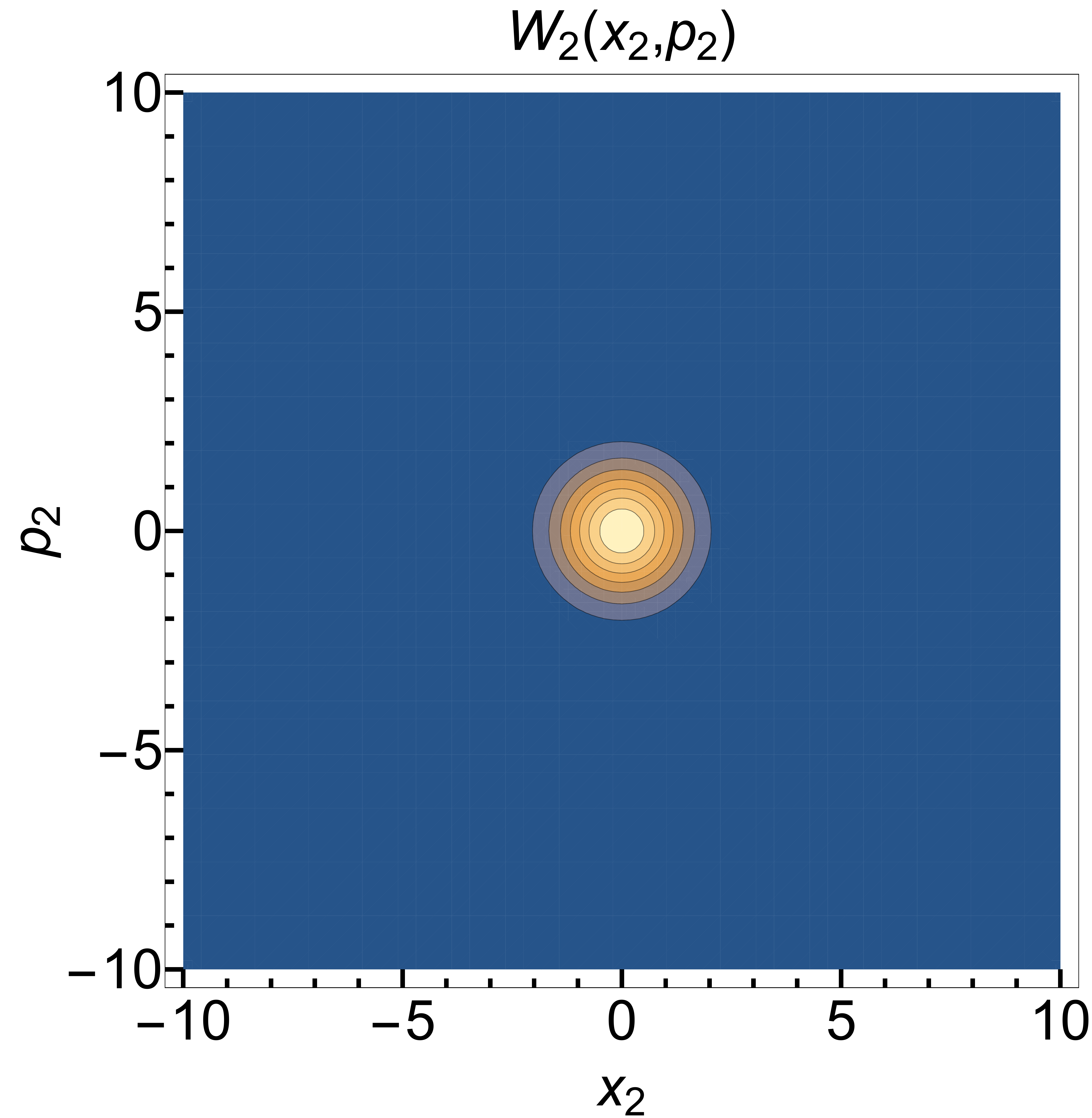}}\\
\subfloat[Channel 1, $\varphi_-=\pi/4$.]{\includegraphics[width=.5\columnwidth]{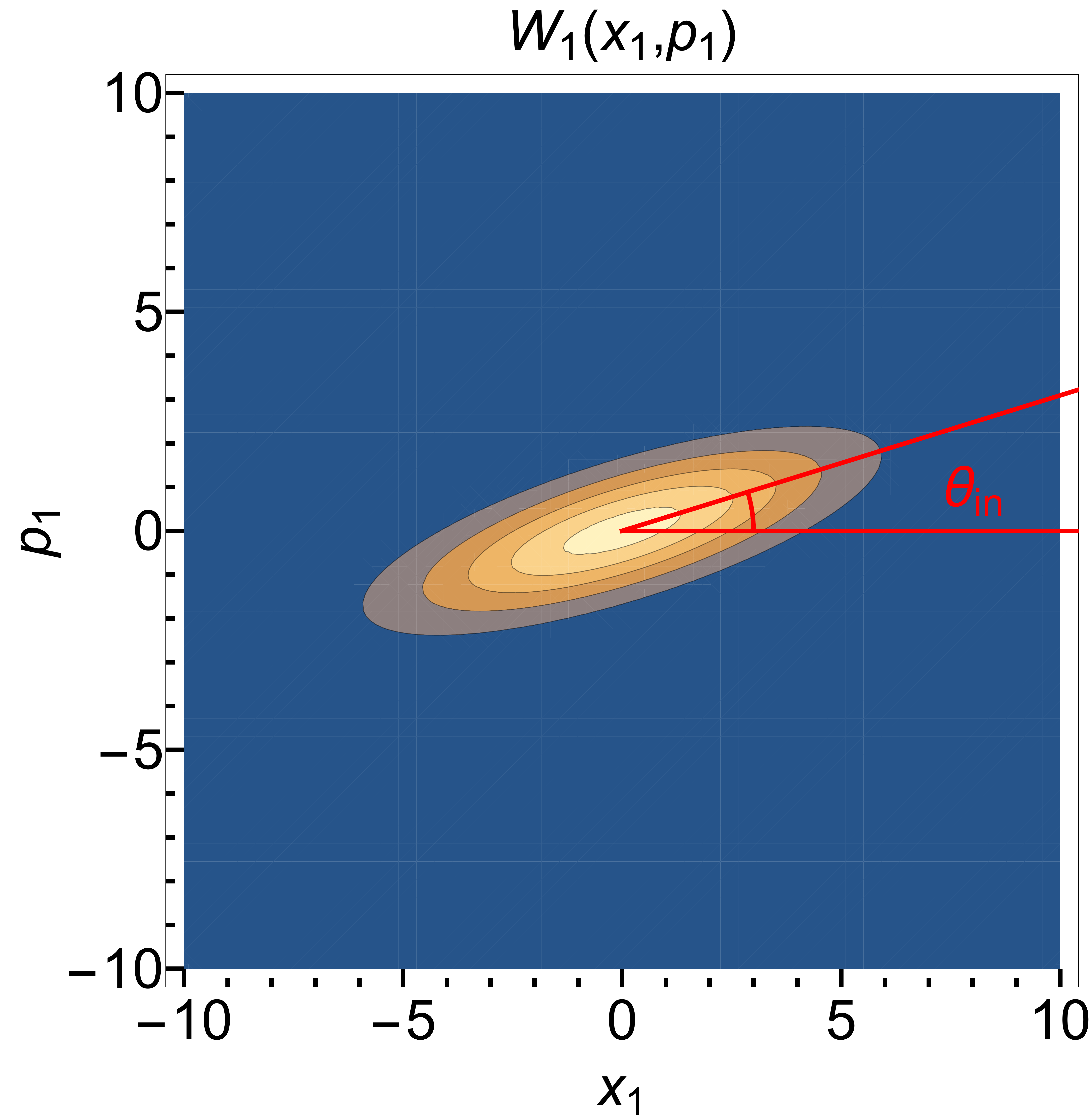}}
\subfloat[Channel 2, $\varphi_-=\pi/4$.]{\includegraphics[width=.5\columnwidth]{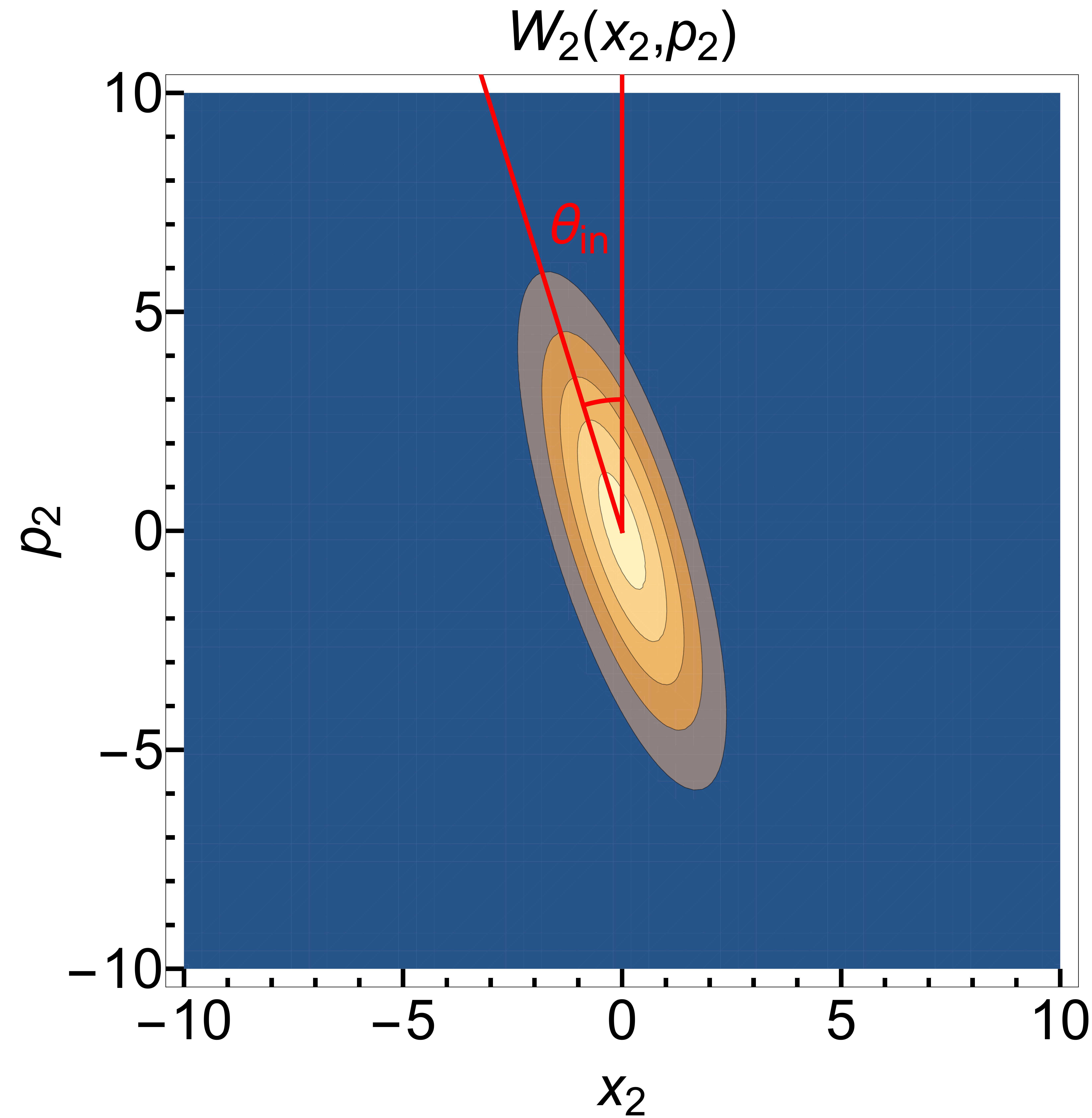}}\\
\subfloat[Channel 1, $\varphi_-=\pi/2$.]{\includegraphics[width=.5\columnwidth]{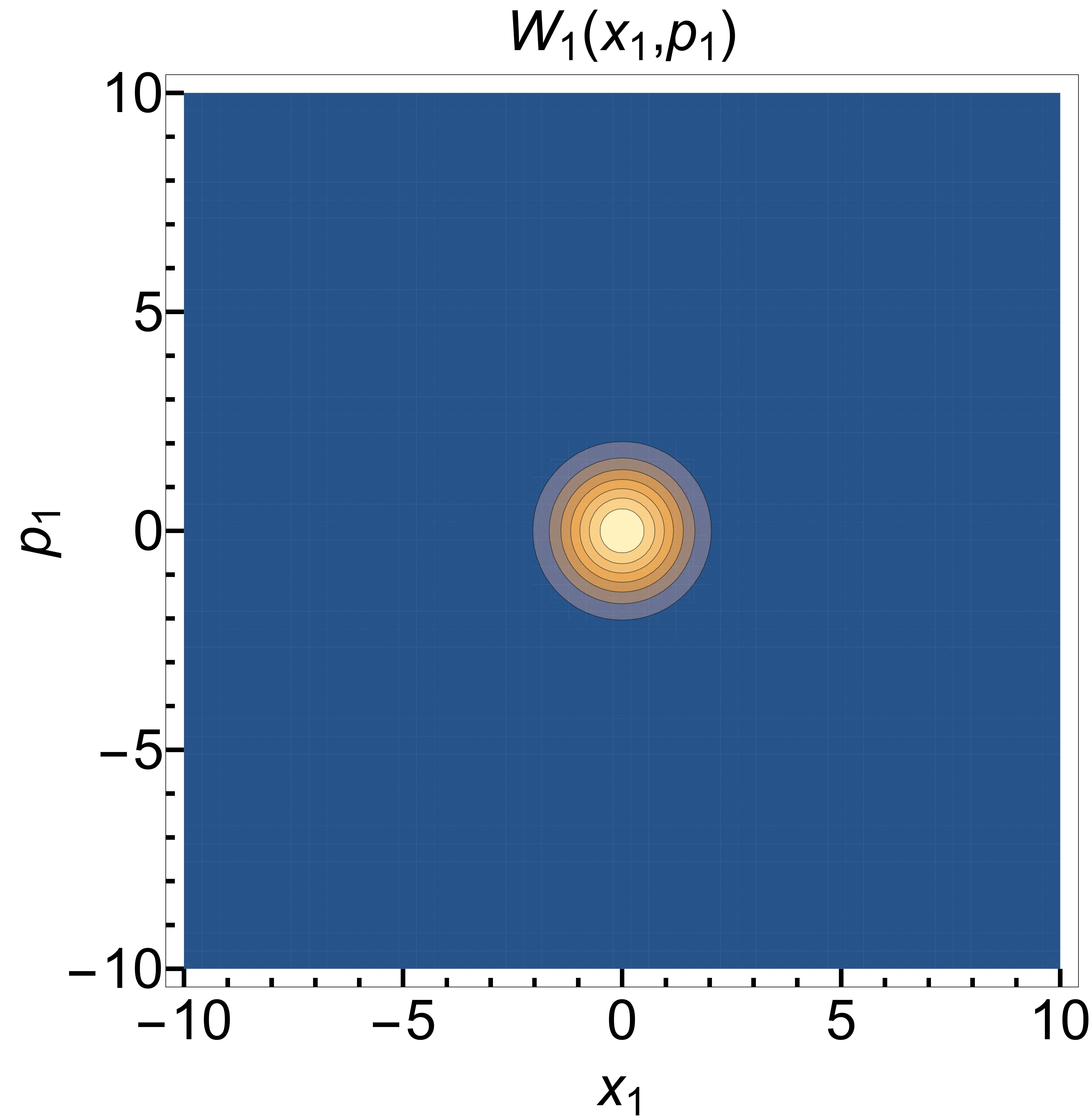}}
\subfloat[Channel 2, $\varphi_-=\pi/2$.]{\includegraphics[width=.5\columnwidth]{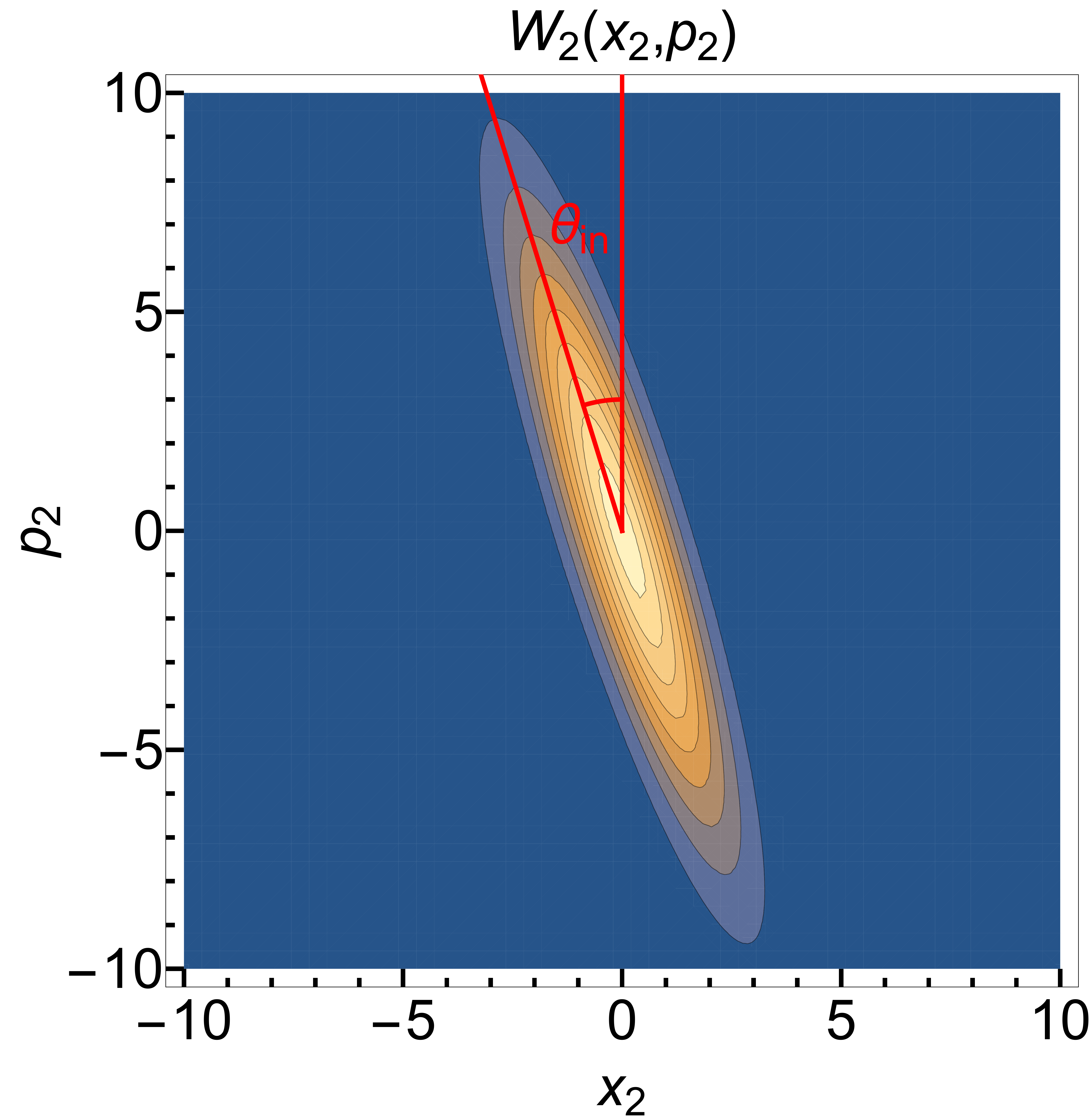}}
\caption{Contour plots of the marginal Wigner functions defined in Eq.~\eqref{eq:reduced_wigners}. The parameter $\varphi_+$ has been set  to zero for simplicity. Notice how at the output of the Mach-Zehnder configuration ($\sigma = \sigma_\mathrm{MZ}$ in Eq.~\eqref{eq:interferometer_cov_transf}) the proportion of photons in each channel, directly related to the squeezing of the ellipses semi-axes as in Eqs.~\eqref{eq:prop_1}-\eqref{eq:prop_2}, shifts from one channel to the other as $\varphi_-$ increases.}
\label{fig:diffphase}
\end{figure}

Initially, the first channel is in a squeezed state and the second one is in the vacuum state corresponding to the covariance matrix $\sigma=\sigma_\mathrm{in}$ in Eq.~\eqref{eq:squeezing_in}: the marginal $W_{2}$ has a circular Gaussian profile, while the marginal $W_{1}$ exhibits an elliptic Gaussian profile with one quadrature below the vacuum level and the other one above it, with the squeezing direction being determined by the angle $\theta_{\mathrm{in}}$ (see Fig.~\ref{fig:diffphase}).
After the photons have traversed the Mach-Zehnder configuration, the profiles undergo the measurement described by Eq.~\eqref{eq:cov_meas}--\eqref{eq:gaussian_overlap}. The parameter $\varphi_-$ controls the squeezing proportions across the two channels, and consequently also what portion of the photons ends up in which channel. The lengths of the ellipses semi-axes can be found to be
\begin{align}
&1+\cos(\varphi_-)^2\Bigl( N \pm\sqrt{ N (1+ N )}\Bigr),
\label{eq:prop_1}  \\
&1+\sin(\varphi_-)^2\Bigl( N \pm\sqrt{ N (1+ N )}\Bigr),
\label{eq:prop_2}
\end{align}
respectively for the first and second channel of the interferometer (see Appendix~\ref{sec:append_00}). 

\begin{figure}
\subfloat[Channel 1, $\varphi_+=0$.]{\includegraphics[width=.5\columnwidth]{A1ii}}
\subfloat[Channel 2, $\varphi_+=0$.]{\includegraphics[width=.5\columnwidth]{A2ii}}\\
\subfloat[Channel 1, $\varphi_+=0.4$.]{\includegraphics[width=.5\columnwidth]{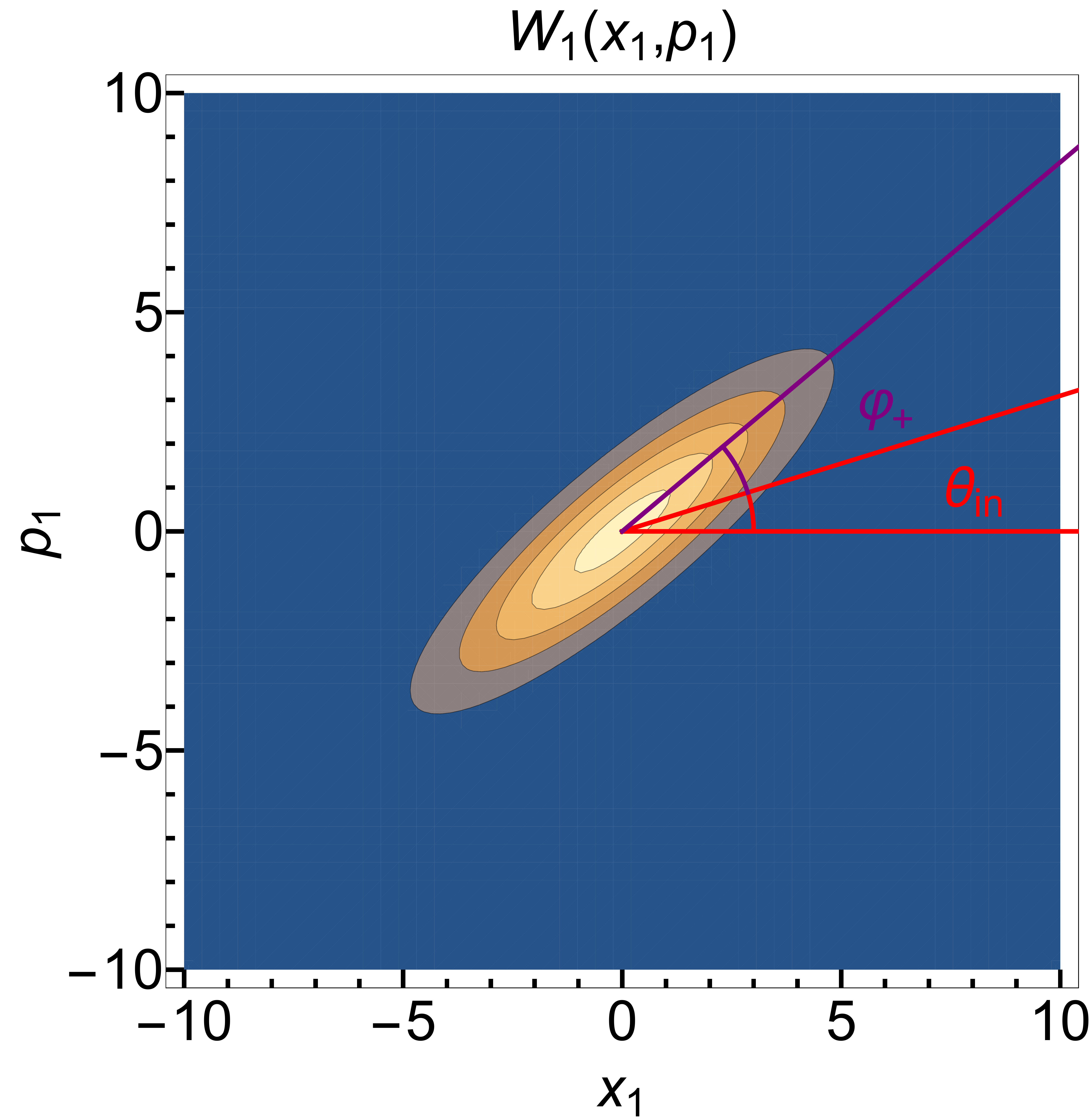}}
\subfloat[Channel 2, $\varphi_+=0.4$.]{\includegraphics[width=.5\columnwidth]{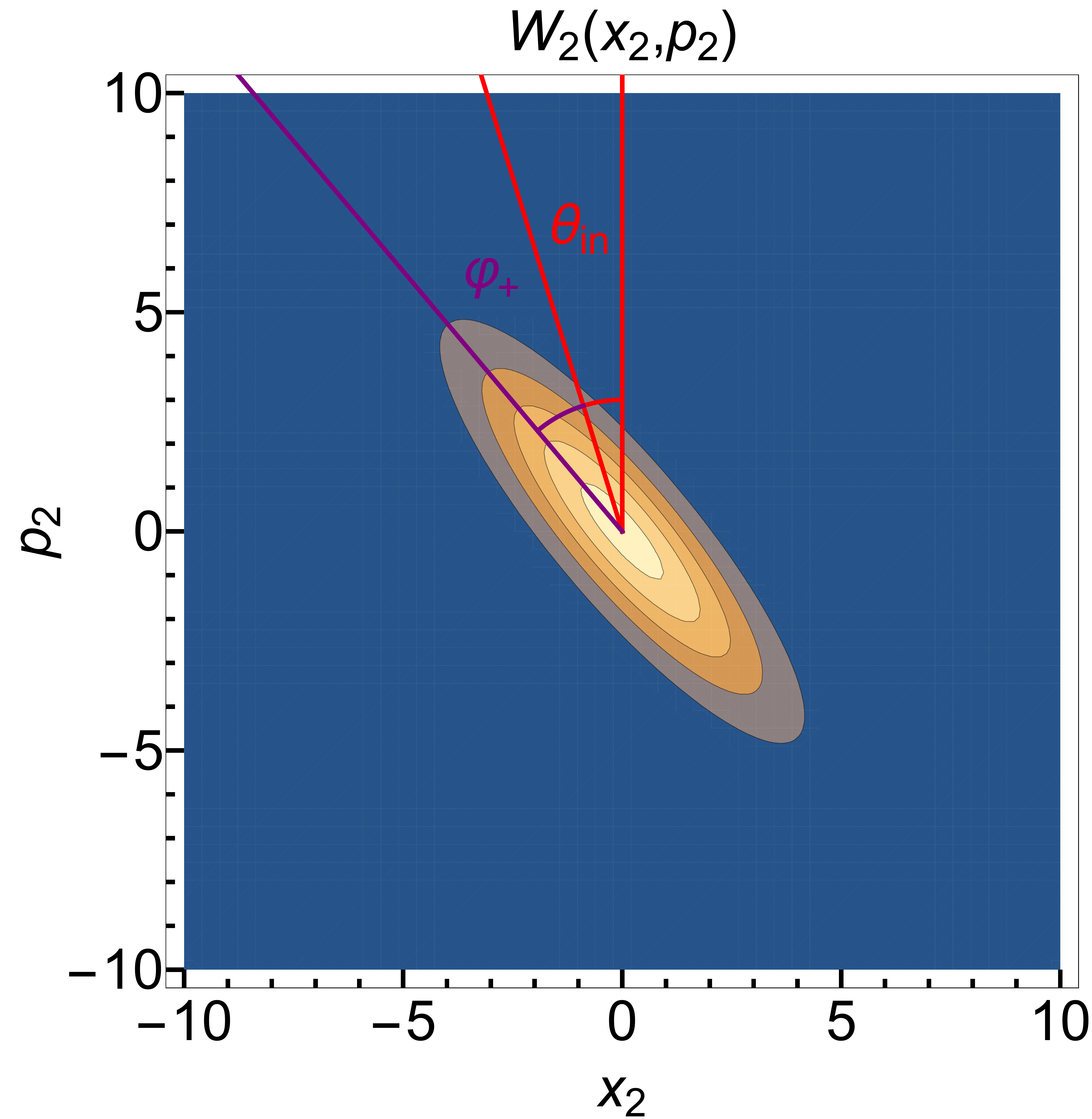}}
\caption{Comparison between the marginal Wigner functions~\eqref{eq:reduced_wigners} for $\sigma=\sigma_\mathrm{MZ}$ in Eq.~\eqref{eq:OMZ} in the case where $\varphi_+ = 0$ (panels a and b) and the ones (panels c and d) rotated in phase space by a non-zero value of $\varphi_+$. The parameter $\varphi_-$ has been set to $\pi/4$ for simplicity.}
\label{fig:sumphase}
\end{figure}

\begin{figure}
 \includegraphics[width=0.8\columnwidth]{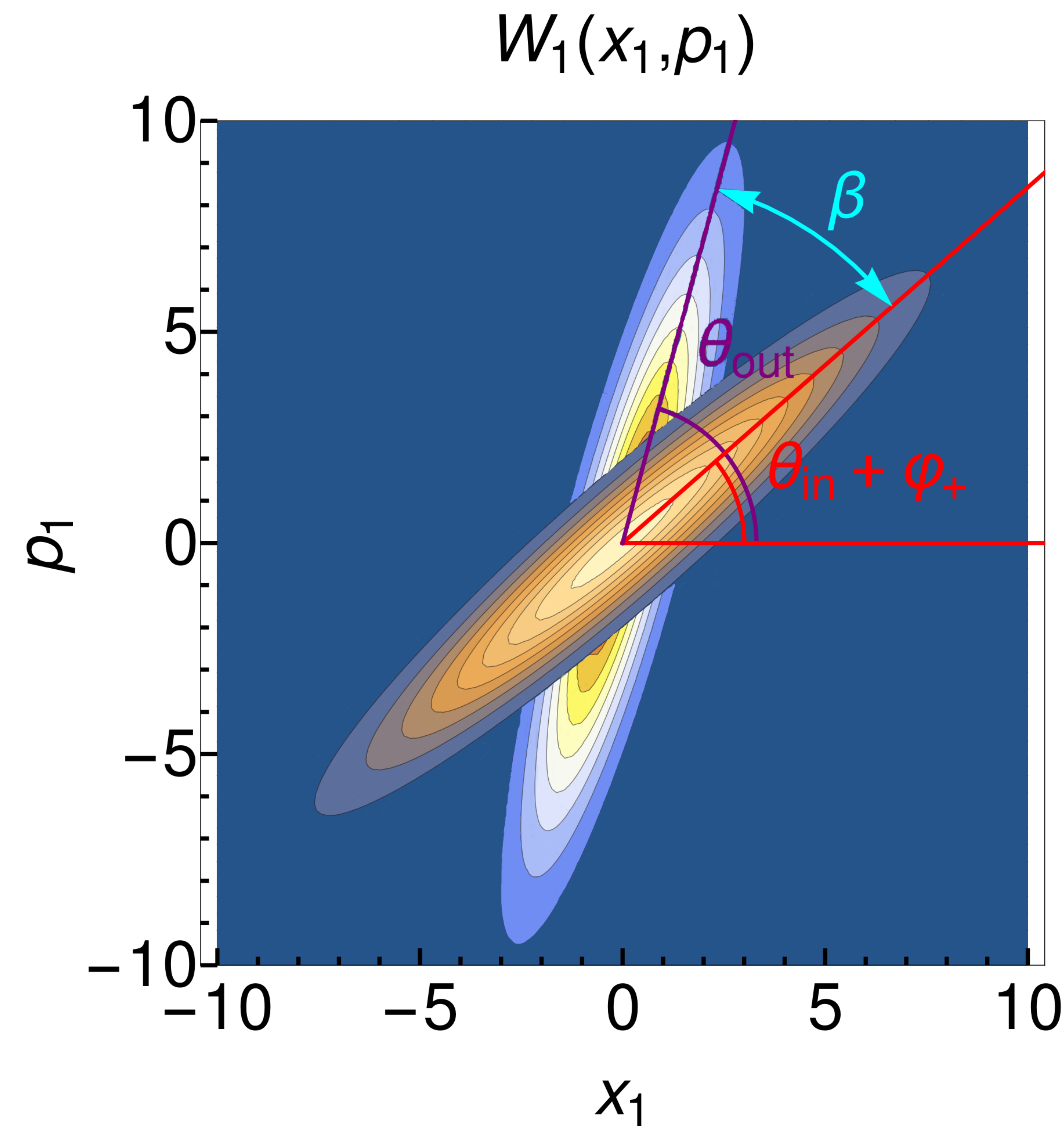}
 \caption{Marginal Wigner functions associated with  the covariance matrices $\sigma_\mathrm{MZ}$ (foreground) and  $\sigma_\mathrm{out}$ (background), defined in Eq.~\eqref{eq:interferometer_cov_transf} and~\eqref{eq:cov_meas}, respectively. The relative angle is $\beta=\varphi_+ + \theta_\mathrm{in} - \theta_\mathrm{out}.$ }
 \label{fig:overlap}
\end{figure}

The effect of the parameter $\varphi_+$, often ignored, is to rotate both profiles by an additional angle of $\varphi_+$ in phase space, through the matrix $e^{-i\varphi_+ \sigma_y}$ in Eq.~\eqref{eq:OMZ} (see Fig.~\ref{fig:sumphase}).
As we shall shortly see, $\varphi_+$ has a physical effect on the outcome of the measurement and thus allows us to estimate the average phase of the Mach-Zehnder. Indeed, when the state is anti-squeezed with the operation $\hat{S}_1^\dag(z_\mathrm{out})$ and then projected onto the vacuum, the detection probability~\eqref{eq:gaussian_overlap}  is given by the overlap between the Gaussian Wigner function $W_{\sigma_\mathrm{MZ}}$ associated with the covariance matrix $\sigma_\mathrm{MZ}$ in Eq.~\eqref{eq:interferometer_cov_transf}, whose marginals have a total phase $\varphi_++\theta_\mathrm{in}$, and the Gaussian Wigner function $W_{\sigma_\mathrm{out}}$ whose marginal $W_1$ at the output channel 1 is rotated in phase space by the squeezing angle $\theta_\mathrm{out}$ associated with the covariance matrix~\eqref{eq:cov_meas}.
As depicted in Fig.~\ref{fig:overlap}, the projection overlap depends on the relative phase 
\begin{equation}
\beta=\varphi_++\theta_\mathrm{in}-\theta_\mathrm{out}
\label{eq:betadef}
\end{equation}
 between the two Wigner functions, as well as the fraction of photons in the first output channel, and thus on $\varphi_-$. 

The probability for ideal detectors to click in the scheme described above is $1-P$, where $P$
is the overlap~\eqref{eq:gaussian_overlap} between the Wigner functions in Eq.~\eqref{eq:wigner_gaussian} with the covariance matrices in Eq.~\eqref{eq:interferometer_cov_transf} and~\eqref{eq:cov_meas}.
Intuitively, we expect this overlap to be maximal when the ellipse associated with $\sigma_\mathrm{MZ}$ is squeezed in the same direction as $\sigma_\mathrm{out}$, i.e. when $\theta_\mathrm{in}+\varphi_+ = \theta_\mathrm{out}$ (see Fig.~\ref{fig:overlap}). 

\begin{figure}
\begin{center}
\subfloat[$ N =4$, $\eta=1$.]{\includegraphics[width=.7\columnwidth]{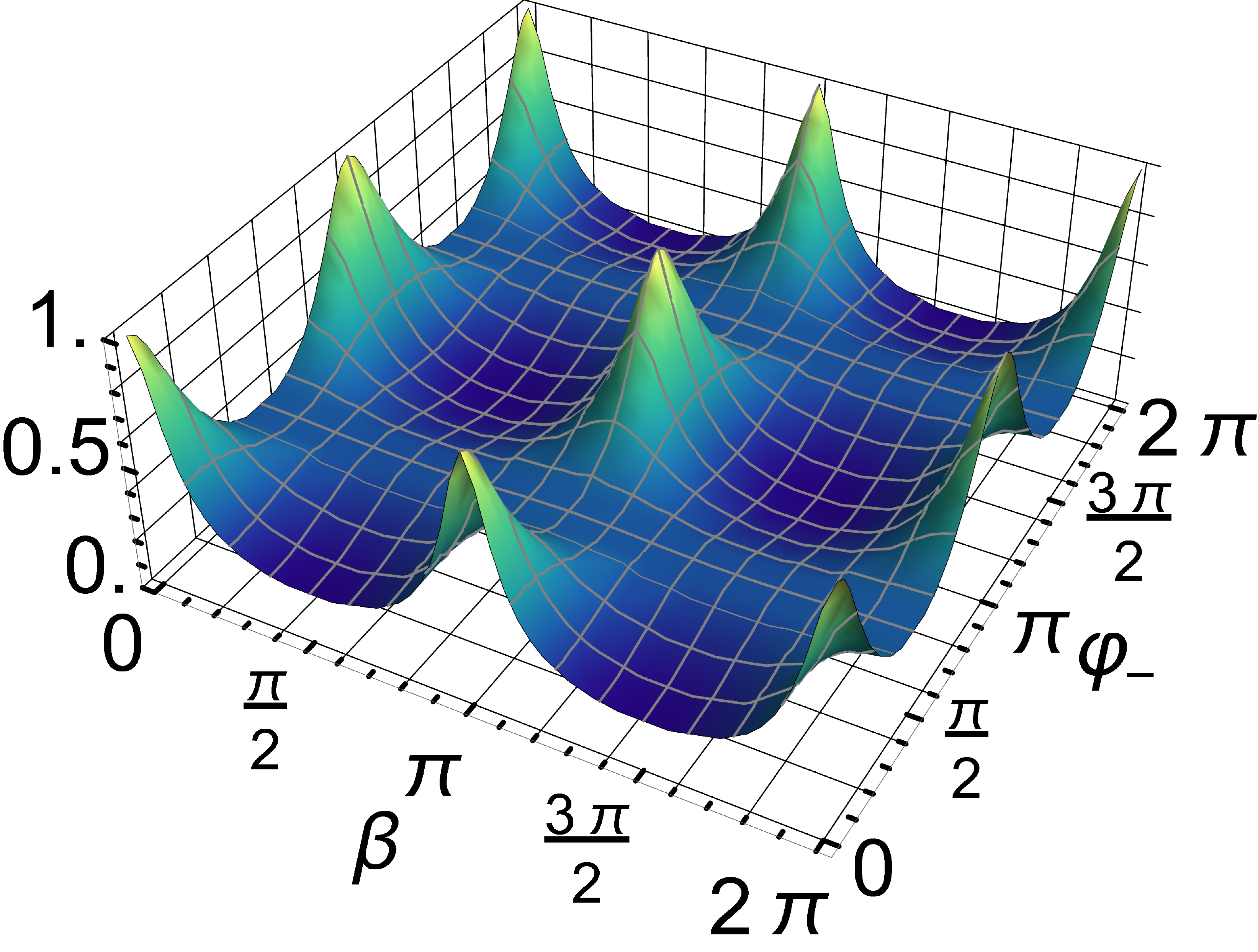}}\\
\subfloat[$ N =20$, $\eta=1$.]{\includegraphics[width=.7\columnwidth]{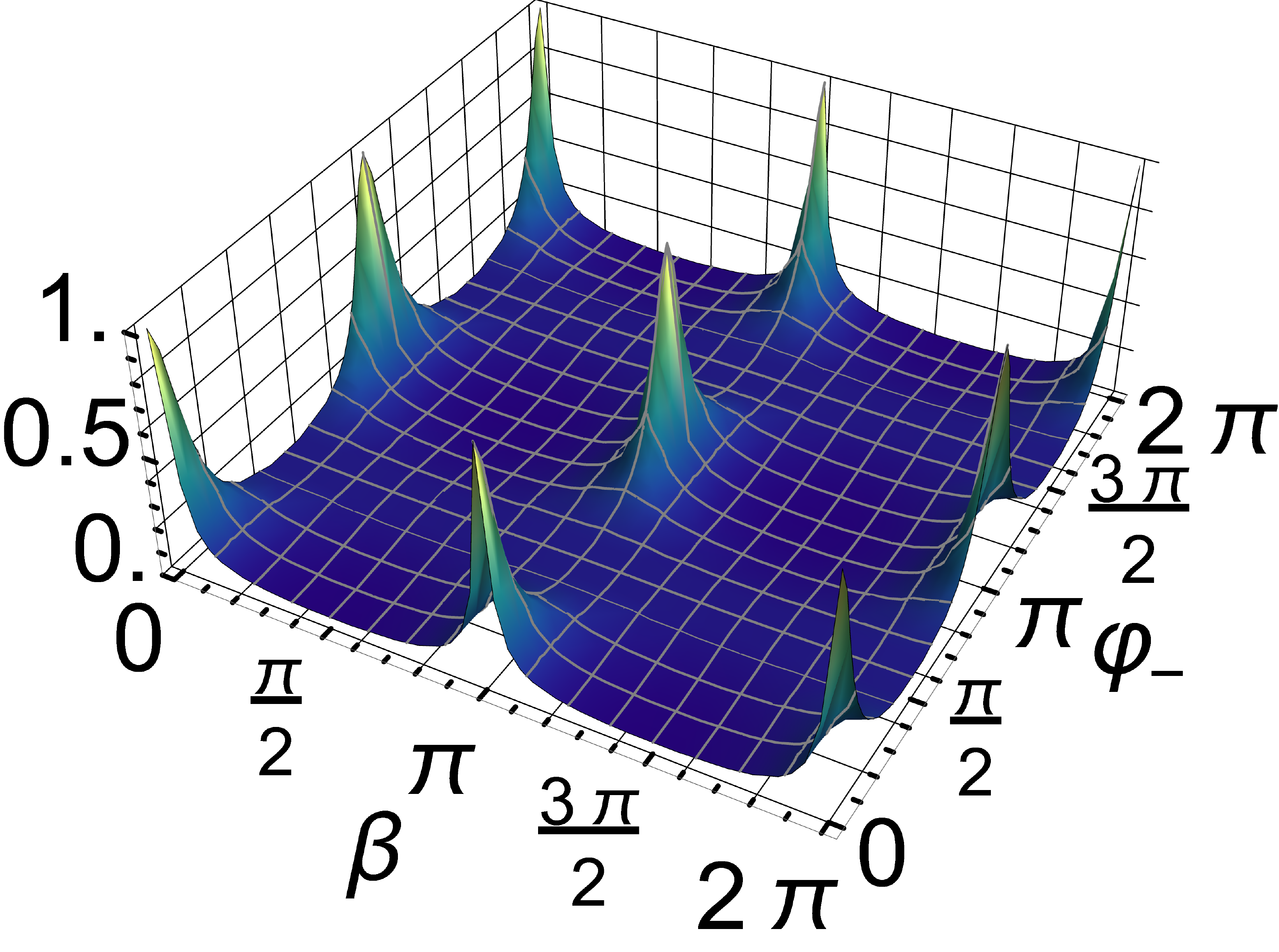}}\\
\subfloat[$ N =4$, $\eta=0.2$.]{\includegraphics[width=.7\columnwidth]{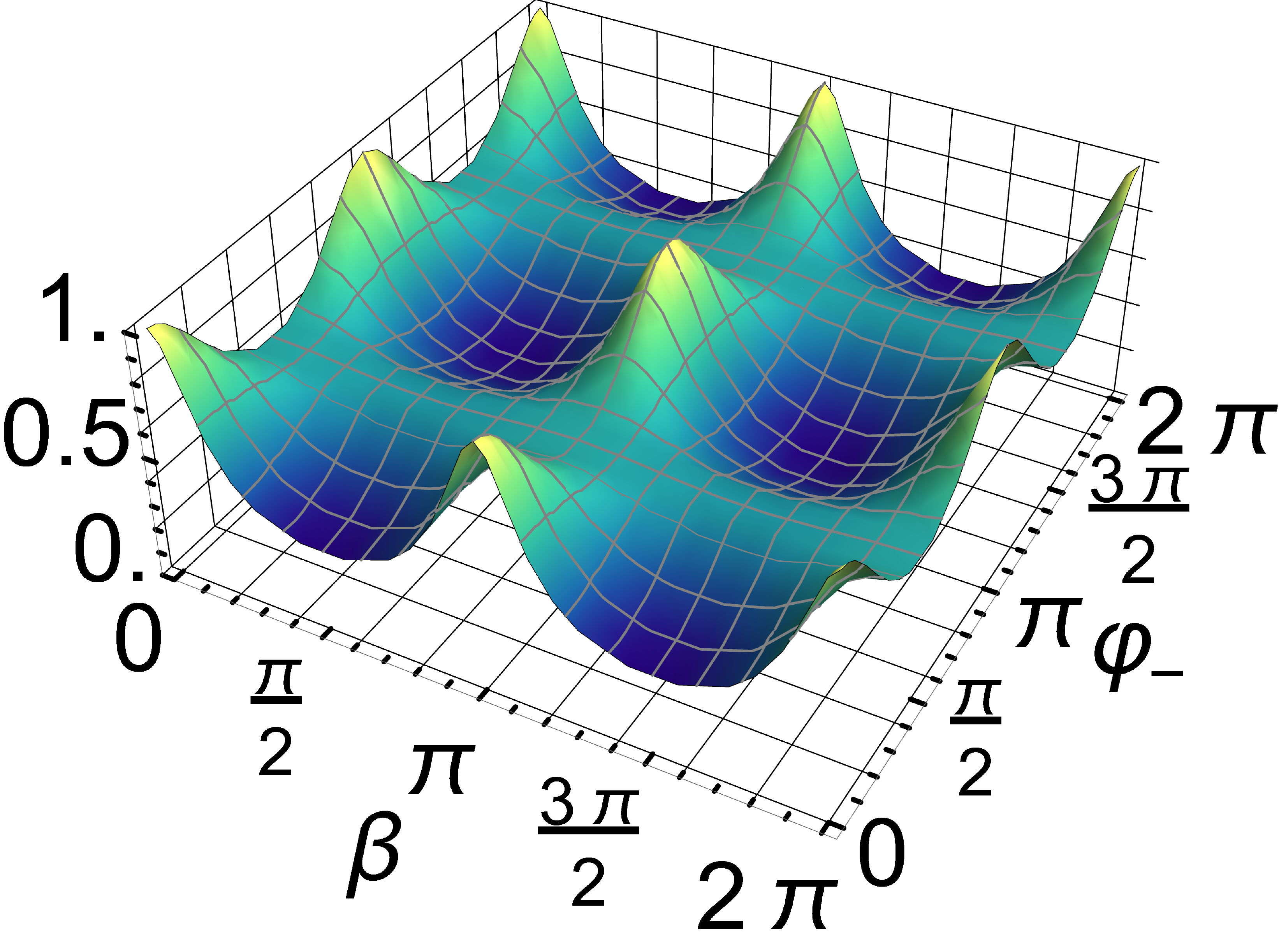}}\\
\subfloat[$ N =6.667$, $\eta=0.2$.]{\includegraphics[width=.7\columnwidth]{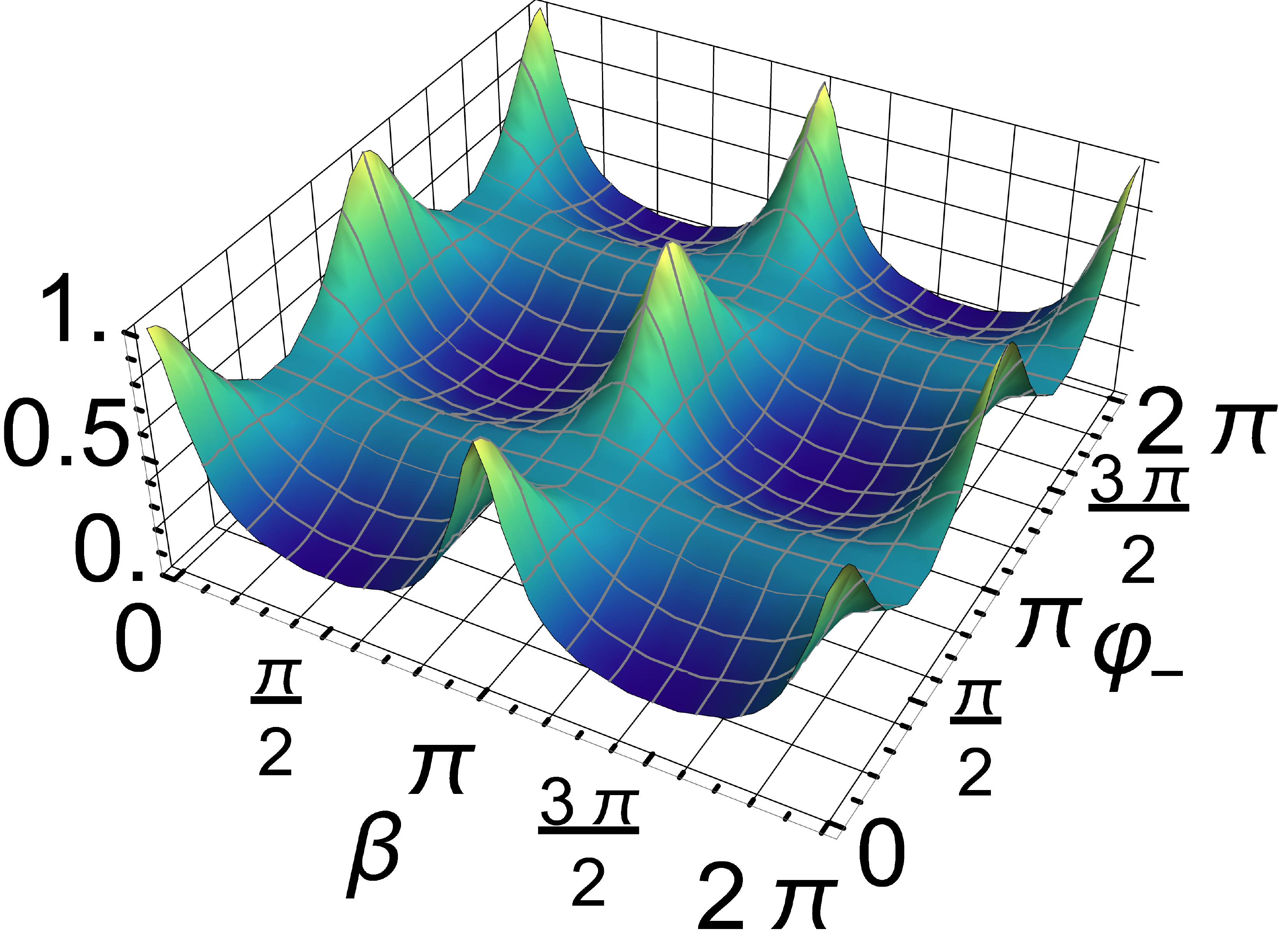}}
\end{center}
\caption{Detection probability~\eqref{eq:detection_prob} as a function of $\beta=\frac{1}{2}(\varphi_1+\varphi_2)+\theta_\mathrm{in}-\theta_\mathrm{out}$ and $\varphi_-=\frac{1}{2}(\varphi_1-\varphi_2)$, for various mean photon numbers $ N $ and quantum efficiencies $\eta$. The maxima of $P$ become more and more localised at the increasing of $ N $, and concentrate way more quickly on the $\beta$-direction than they do on the $\varphi_-$-direction. The spread caused by a non-unit quantum efficiency is compensated by dividing $ N $ by the factor $\sqrt{\tilde{\eta}} = \sqrt{\eta(2-\eta)}$, as is evident comparing  the panels a and d in the case of $\eta = 0.2$.}
\label{fig:prob}
\end{figure}

To account for a detector quantum efficiency $\eta$ (with $0<\eta\leq 1$), we imagine that the (ideal) detectors are preceded by a fictitious beam-splitter with reflectivity $\eta$. This modifies Eq.~\eqref{eq:gaussian_overlap} as follows: 
\begin{equation}
 P = \det\big(\eta\,\sigma_\mathrm{MZ}+(2-\eta)\sigma_\mathrm{out}\big)^{-1/2},
 \label{eq:14}
\end{equation}
as shown in Appendices~\ref{sec:append_1} and~\ref{sec:append_2}. Using the expressions~\eqref{eq:interferometer_cov_transf} and~\eqref{eq:cov_meas} of the covariance matrices, the detection probability~\eqref{eq:14} reads
\begin{widetext}
 \begin{align}
P(\beta,\varphi_-) = \bigg\{1 + \tilde{\eta}\bigg[2 N  + \Big(2\cos(\varphi_-)^2 + \tilde{\eta} \sin(\varphi_-)^4\Big) N ^2 
- 2\cos(\varphi_-)^2\cos(2\beta)\, N (1+ N )\bigg]\bigg\}^{-1/2}, 
\label{eq:detection_prob}
\end{align}
\end{widetext}
with 
\begin{equation}
\label{eq:tildeta}
\widetilde{\eta} = \eta (2-\eta),
\end{equation}
and the relative phase $\beta$ given in Eq.~\eqref{eq:betadef}. Plots of $P$ as a function of $\beta$ and $\varphi_-$ are provided in Fig.~\ref{fig:prob} for different values of the mean photon number~$N$ and the detector quantum efficiency~$\eta$.

From Eq.~\eqref{eq:detection_prob} and Fig.~\ref{fig:prob} the periodicity of $P$ with period $\pi$ in both variables is evident, hence we may restrict our attention to a fundamental domain, such as $|\beta|\le\pi/2$, $|\varphi_-|\le\pi/2$. The detection probability $P$ is maximal for $(\beta,\varphi_-)=(h,k)\,\pi$, with integer $h$ and $k$. As can be seen from Fig.~\ref{fig:prob}, these peaks become more and more localised as $ N $ increases, with a variation on the $\beta$-axis significantly more stark than on the $\varphi_-$-axis. A non-unit quantum efficiency instead spreads out these peaks, an effect which is quickly compensated by a moderate increase of $ N $ by a factor which later we will show to be equal to $\sqrt{\widetilde{\eta}}$. Furthermore, notice how both $\varphi_-$ and $\beta$ affect the detection probability: the relative angle $\beta$ between the squeezed ellipses (corresponding to $W_{\sigma_\mathrm{MZ}}$ and $W_{\sigma_\mathrm{out}}$ in Eq.~\eqref{eq:gaussian_overlap}) has a major physical effect, allowing both the global phase of the Mach-Zehnder $\varphi_+$ and the relative phase of the squeezers $\theta_\mathrm{in}-\theta_\mathrm{out}$ to be estimated, once the other one is known. Thus, assuming that only the relative phase in a Mach-Zehnder interferometer bears physical significance is not correct in general: for a discussion of the cases when this is actually correct see Ref.~\cite{gatto19-2}. In fact, we will see it is impossible to estimate $\beta$ (and hence $\varphi_+$) without any knowledge of $\varphi_-$, and viceversa. Of course, in practice, any laser source must be phase-locked with respect to a reference phase. This is reflected in the dependence of $P$ from the relative phase \textit{of the squeezers}, $\theta_{\mathrm{in}}-\theta_{\mathrm{out}}$, which shifts the probability plot on the $\beta$-axis.  Analogously, one could estimate $\theta_\mathrm{in} - \theta_\mathrm{out}$ when the known reference phase is the average phase $\varphi_+$ in the network.

\section{Estimation procedure}
Let us now discuss the sensitivity achievable with this estimation scheme. The relative angle of the squeezers---or any other combination of $\varphi_1,\varphi_2,\theta_\mathrm{in}$ and $\theta_\mathrm{out}$ over which the experimentalist has good control---can be used to calibrate the estimation apparatus so that the detectors click at almost every trial, within the desired confidence level. 

\begin{figure}
 \includegraphics[width=0.9\columnwidth]{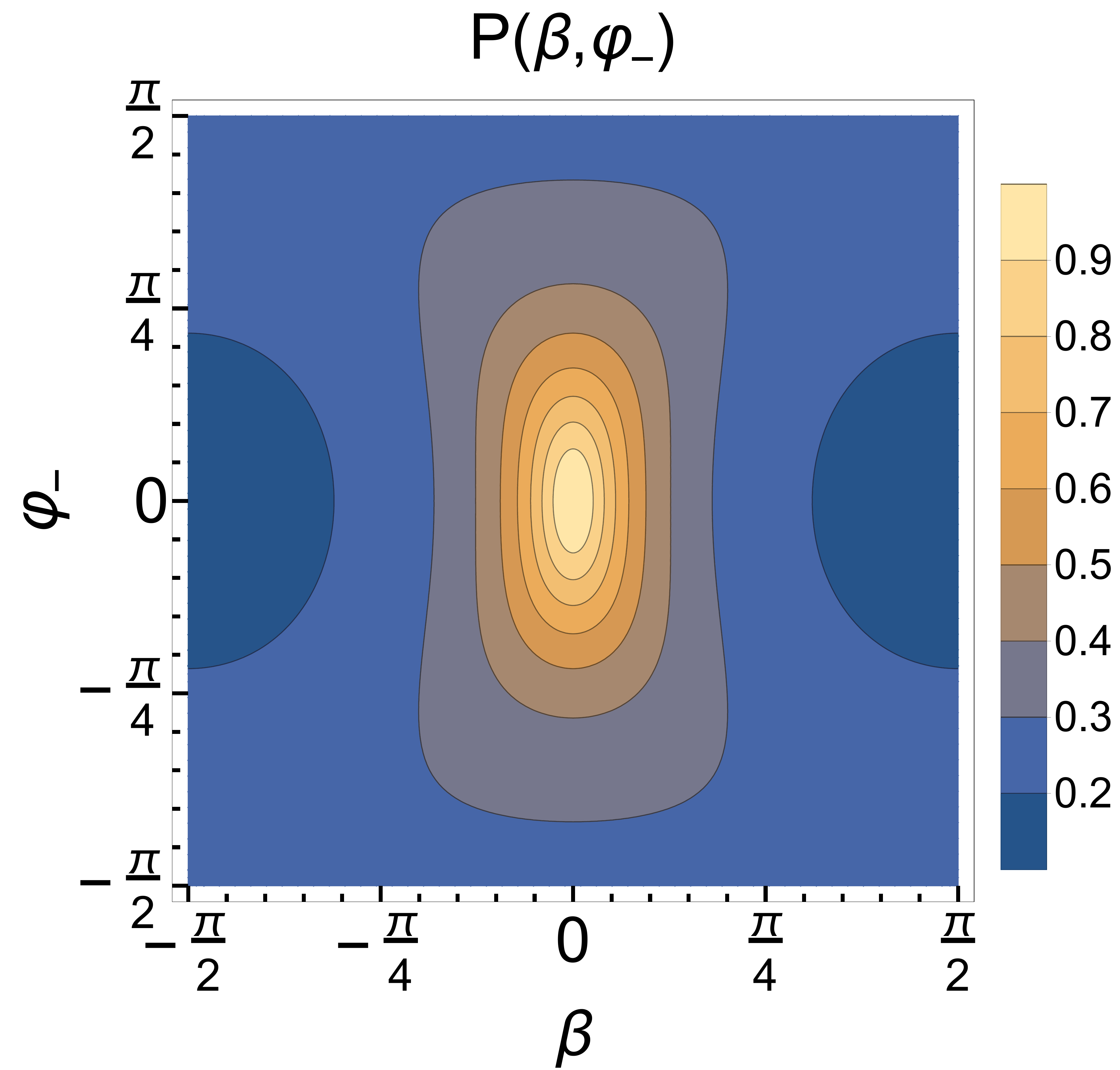}
 \caption{Level curves for the detection probability $P$ given in Eq.~\eqref{eq:detection_prob}. Every couple of parameters corresponding to a point on a given level curve is completely indistinguishable from every other point on the same curve.}
 \label{fig:levels}
\end{figure}

\begin{figure}
\begin{center}
 \includegraphics[width=0.85\columnwidth]{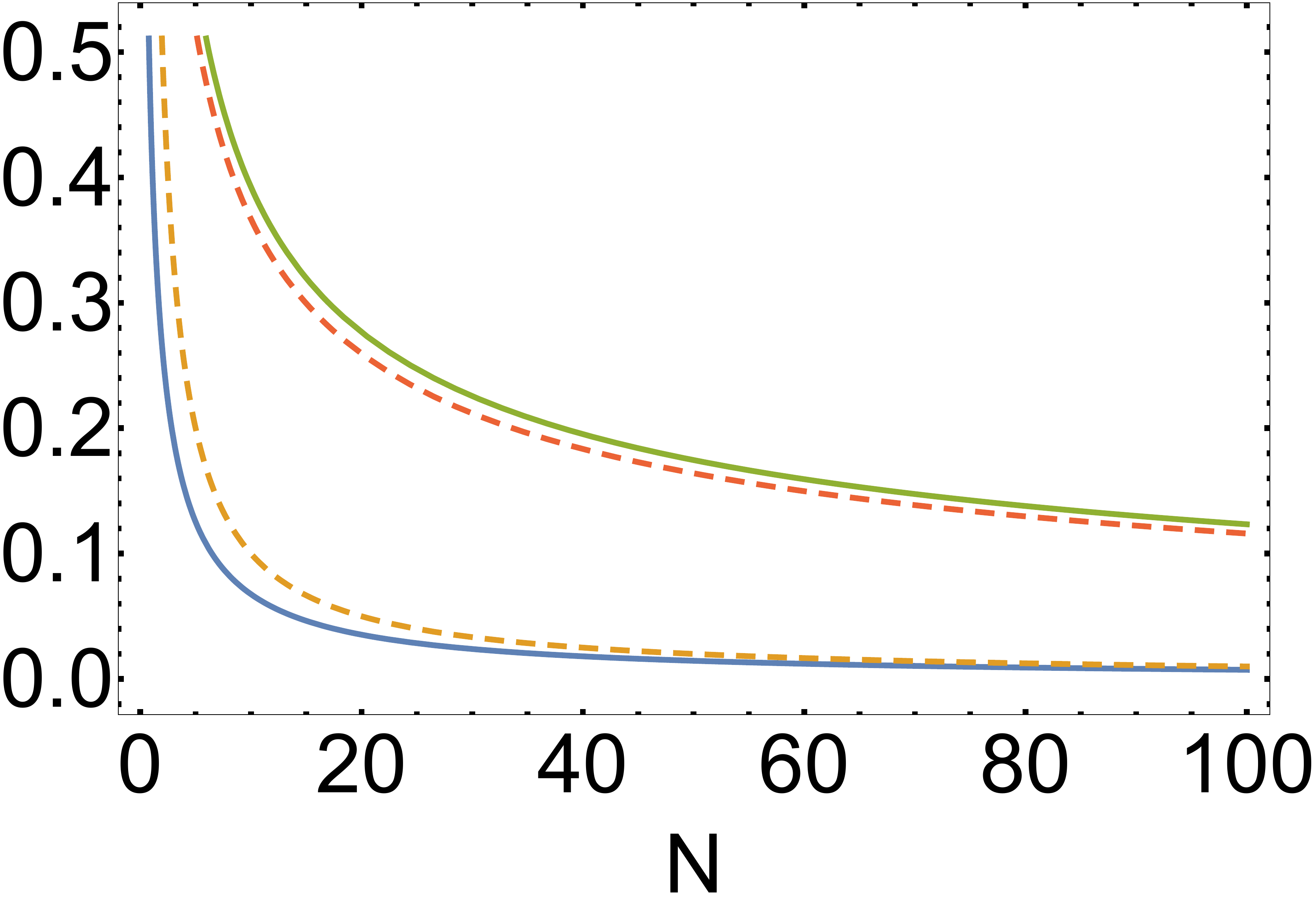}
\end{center}
 \caption{Diameters for the 90\% level curve of the detection probability $P$ as a function of $ N $, at $\eta=1$. The diameter on the $\beta$-direction (bottom solid blue line) is compared with $1/ N $ (bottom dashed yellow line), whereas the diameter on the $\varphi_-$-direction (top solid green line) is compared with $1/\sqrt{ N }$ (top dashed red line).}
 \label{fig:diameters}
\end{figure}

The quantum observable measured at the output of the interferometer is the projector $\hat{\Pi}$ in~\eqref{eq:projector}, and the detection probability is given in Eq.~\eqref{eq:gaussian_overlap}. In order to correctly estimate $\beta$ or $\varphi_-$ we have to deal with the local indistinguishability of the parameters~\cite{stoica01,stoica82,li12,rothenberg71}. This means that for any fixed probability $P_0$, a given couple of parameters $(\beta,\varphi_-)$ is indistinguishable from every other point on the curve $P(\beta,\varphi_-)=P_0$, as all possible outcomes of the measurement of $\hat{\Pi}$ have the same probabilities in either case (see Fig.~\ref{fig:levels}).
The ambiguity on $\beta$ is removed once the value of $\varphi_-$ is known, but clearly in practice $\varphi_-$ cannot be known with arbitrary precision. In the previous section we have observed that the peaks concentrate more quickly on the $\beta$-direction than on the $\varphi_-$-direction: more precisely, the horizontal diameter of the level curve corresponding to a fixed probability $P_0$ can be determined directly from Eq.~\eqref{eq:detection_prob}, and reads (see Appendix~\ref{sec:append_3} for details)
\begin{equation}
\beta_* = \arcsin\sqrt{\frac{1-P_0^2}{4\tilde{\eta} N (1+ N )P_0^2}}.
\label{eq:boh_4}
\end{equation}
Similarly, the vertical diameter can be found to be
\begin{equation}
\varphi_* = \arcsin\sqrt{\frac{\sqrt{P_0^2+\tilde{\eta}(1-P_0^2)} -P_0}{\tilde{\eta} N P_0}}.
\label{eq:boh_5}
\end{equation}
From these expressions, plotted in Fig.~\ref{fig:diameters} versus $N$, it is evident that
 the horizontal diameter $\beta_*$ scales as $1/ N $, whereas the vertical diameter $\varphi_*$ scales as $1/\sqrt{ N }$.
Thus a knowledge of $\varphi_-$ with classical precision will be sufficient to estimate $\beta$ with Heisenberg-limited sensitivity, as the uncertainty in $\varphi_-$ becomes irrelevant for large $ N $. 

It remains to be shown that the Heisenberg limit can be achieved in this way. If the measurement of $\hat{\Pi}$ in~\eqref{eq:projector}
is repeated $n$ times and the outcomes are $x_1,...\,,x_n$, with $x_i=0$ or 1 according to whether the detector clicked or not, the maximum likelihood estimator $\widetilde{\beta}$ is defined implicitly by the equation
\begin{equation}
P(\widetilde{\beta},\varphi_-) = \frac{1}{n}\sum_{i=1}^nx_i.
\label{eq:MLE_def_b}
\end{equation}
For large $n$, the variance of $\widetilde{\beta}$ can be approximated as
\begin{align}
\mathrm{Var}[\widetilde{\beta}] &\simeq \frac{\langle\hat{\Pi^2}\rangle -\langle\hat{\Pi}\rangle^2}{n(\partial\langle\hat{\Pi}\rangle/\partial\beta)^2}  \nonumber\\
&=\frac{P(\beta,\varphi_-)\big(1-P(\beta,\varphi_-)\big)}{n(\partial P/\partial\beta)^2}.
\label{eq:sensitivity}
\end{align} 
When all the terms in expression~\eqref{eq:sensitivity} are non-zero, since from~\eqref{eq:detection_prob} $P\propto 1/ N$, one would have that~\eqref{eq:sensitivity} scales as $\mathrm{Var}[\widetilde{\beta}]  \propto N$. Thus we cannot expect a good metrological scaling for generic values of $\beta$ and $\varphi_-$. However in correspondence of any of the maxima in Fig.~\ref{fig:prob} we have $P=1$ independently of $ N $, and both the numerator and the denominator of Eq.~\eqref{eq:sensitivity} vanishes; we will momentarily show that  there is Heisenberg scaling in this case. 

\begin{figure}
\begin{center}
\subfloat[$1/ N ^2\mathrm{Var}(\widetilde{\beta})$, $ N =2$.]{\includegraphics[width=.65\columnwidth]{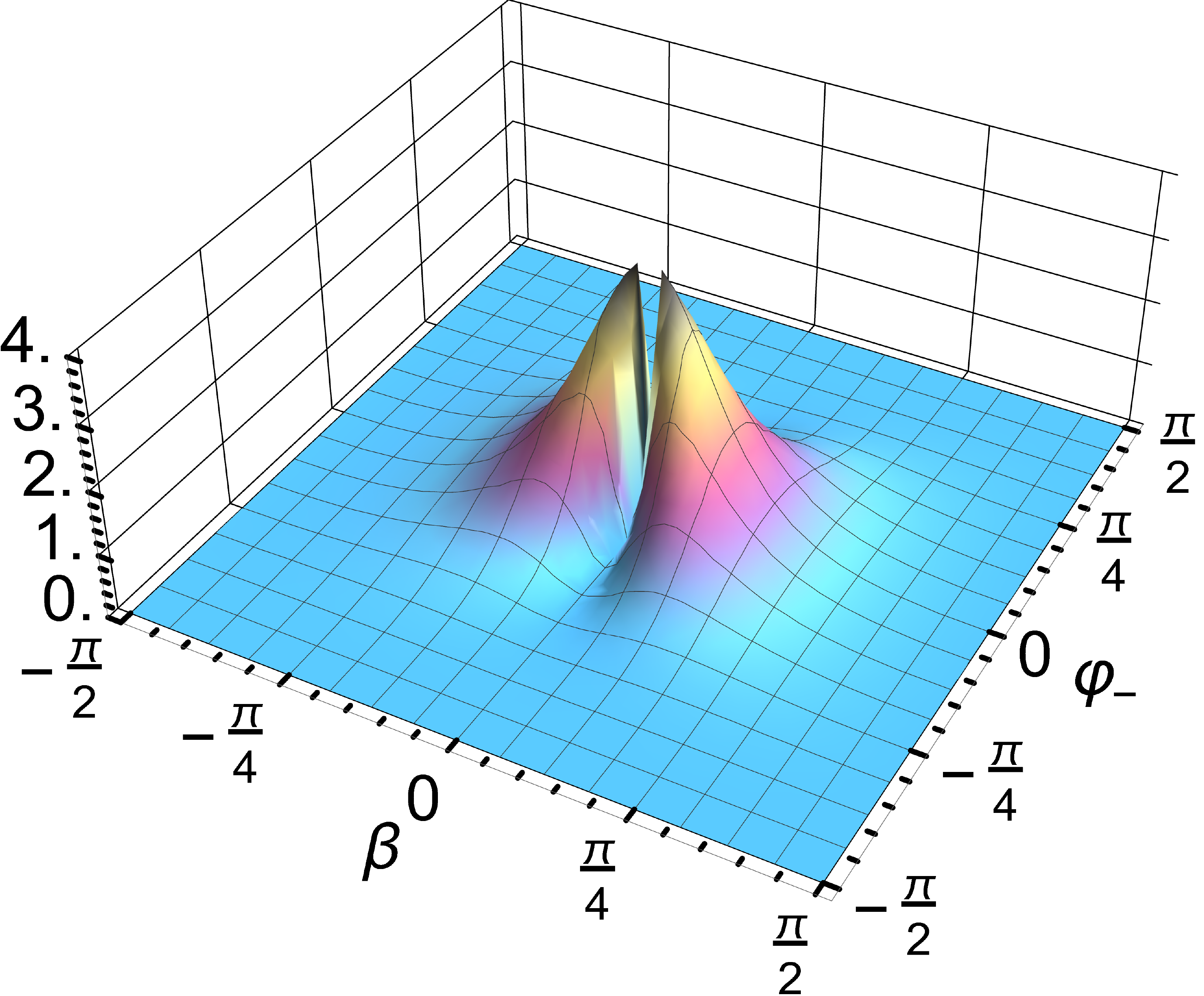}}\\
\subfloat[$1/ N ^2\mathrm{Var}(\widetilde{\beta})$, $ N =20$.]{\includegraphics[width=.65\columnwidth]{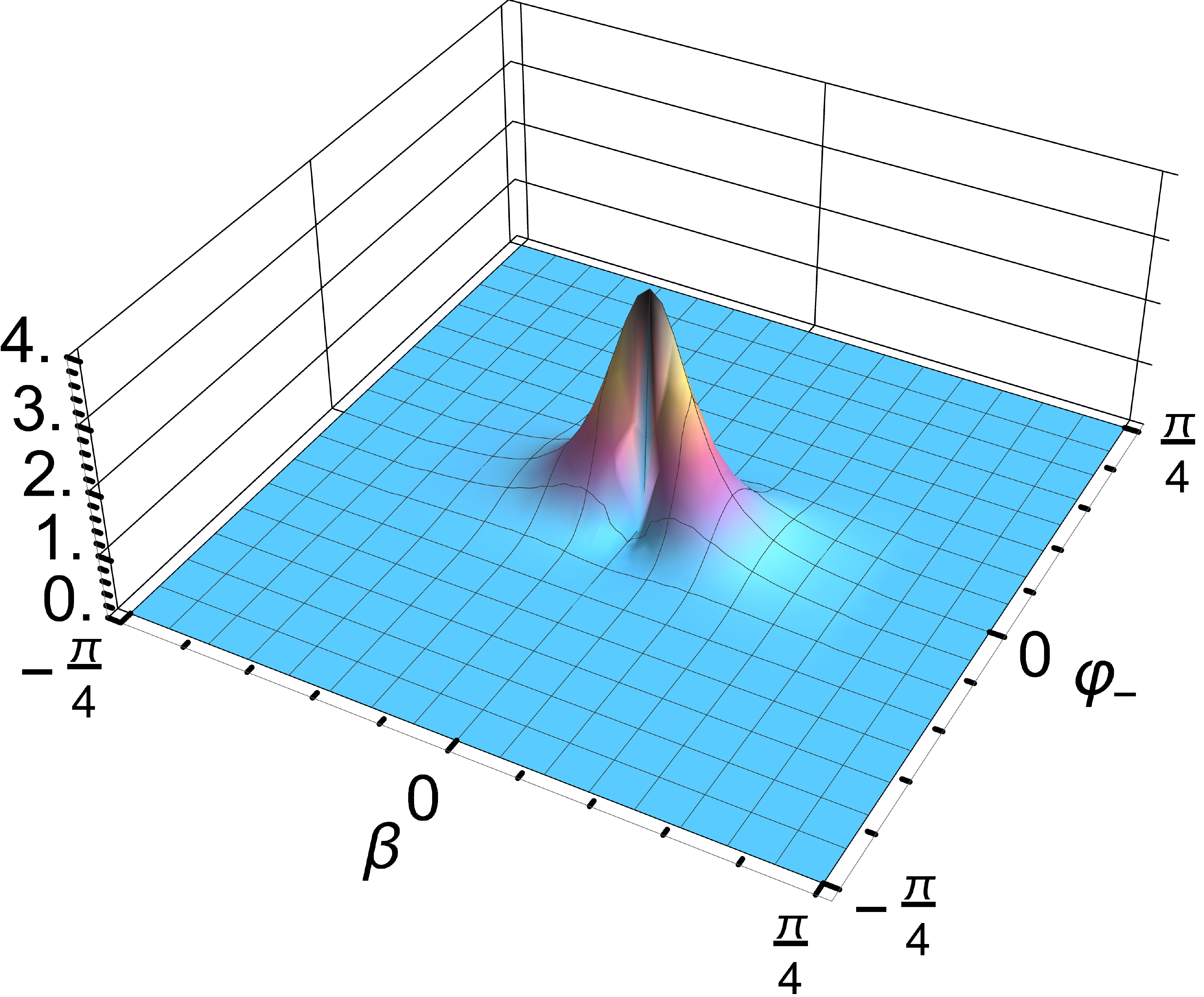}}\\
\subfloat[$1/ N \mathrm{Var}(\widetilde{\varphi_-})$, $ N =2$.]{\includegraphics[width=.65\columnwidth]{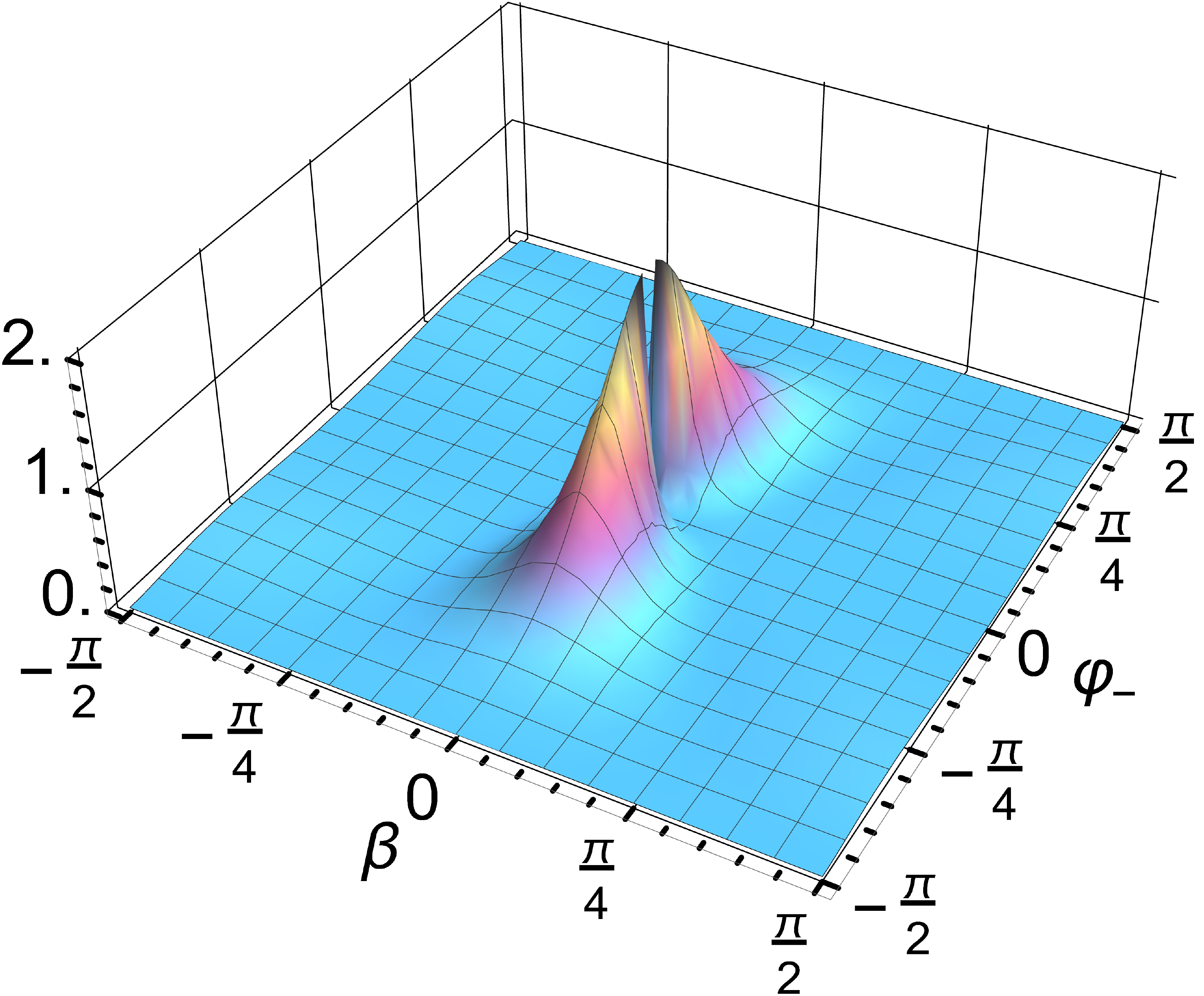}}\\
\subfloat[$1/ N \mathrm{Var}(\widetilde{\varphi_-})$,~$ N =20$.]{\includegraphics[width=.65\columnwidth]{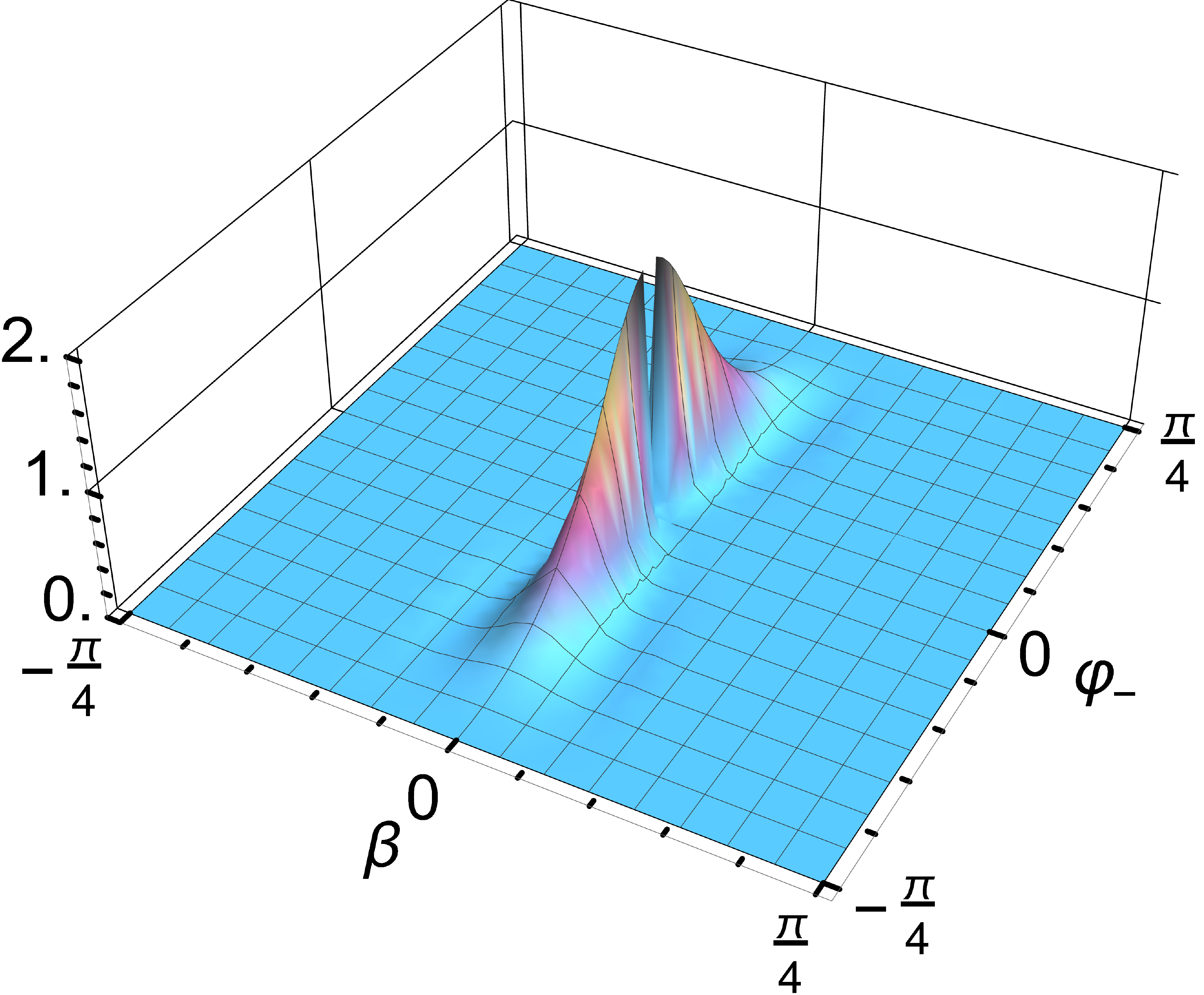}}
\end{center}
\caption{Plots of the rescaled inverse variances of $\widetilde{\beta}$ (panels a and b), and $\widetilde{\varphi_-}$ (panels c and d), as $ N $ increases. If the variance of $\widetilde{\beta}$ is rescaled by $ N ^2$, and the variance of $\widetilde{\varphi_-}$ by $ N $, there is a neighbourhood of the origin where they both stay constant. The jump discontinuity at the origin is a consequence of the local indistinguishability of the parameters.}
\label{fig:rescaled_vars}
\end{figure}

A plot of the rescaled sensitivity $1/ N ^2\mathrm{Var}[\widetilde{\beta}]$, which we report in Fig.~\ref{fig:rescaled_vars}, shows that there is in fact a neighbourhood of the origin which stays essentially constant as $ N $ increases, indicating there  is indeed Heisenberg scaling in this region. 
Interestingly, this suggests constructive quantum interference is a necessary metrological resource for reaching the Heisenberg limit. 
The jump discontinuity at the origin is again a consequence of the local indisinguishability of the parameters: performing the estimation once $\varphi_-$ is known corresponds to taking a \textit{section} of Fig.~\ref{fig:rescaled_vars} at fixed $\varphi_-$, and each of these sections is well-defined and free from singularities. 

Therefore we require $\varphi_-\simeq h\pi$, $h$ integer, with classical precision. Besides making sure the sensitivity~\eqref{eq:sensitivity} stays close to its maximal value, where Heisenberg scaling is expected, the condition $\varphi_-\simeq h\pi$ entails that all photons end up in the first channel, offering the practical advantage of having to place only one detector at the output of the interferometer. 

Moreover, our discussion indicates the size of the Heisenberg-limited region about a maximum scales as $1/ N $ in the $\beta$-direction. Taking these fact into account, we can expand expression~\eqref{eq:sensitivity} around its maxima, that is
$\beta=k\pi+\delta\beta$ and $\varphi_-=h\pi+\delta\varphi$, with $|\delta\beta|\le\beta_*$ and $|\delta\varphi|\le\varphi_*$, and obtain
\begin{widetext}
\begin{align}
\mathrm{Var}[\widetilde{\beta}] &= \frac{P(\beta,\varphi_-)\big(1-P(\beta,\,\varphi_-)\big)}{n(\partial P/\partial\beta)^2}\bigg|_{\beta=k\pi+\delta\beta,\varphi_-=h\pi+\delta\varphi} 
\nonumber\\
&= \frac{1}{32\tilde{\eta}
n N ^2}   +\frac{\big(1+ \tilde{\eta}(2\delta\beta^2 N ^2+\delta\varphi^2 N )\big)\big(\sqrt{1+\tilde{\eta}\delta\beta^2 N ^2}-1\big)(2\delta\beta^2 N ^2+\delta\varphi^2 N )}{24\tilde{\eta}^2\delta\beta^2n N ^5} + O\bigg(\frac{1}{ N ^4}\bigg).
\label{eq:asymptotic_sensitivity}
\end{align}
\end{widetext}
Since $\delta\beta = O(1/ N) $ and $\delta\varphi = O(1/\sqrt{ N })$, the first term in Eq.~\eqref{eq:asymptotic_sensitivity}, which scales as $1/ N ^2$, is the dominant one, the second term being of order $1/N^3$. 
Hence, the estimation of $\beta$ can be achieved at the Heisenberg limit, with a statistical error
\begin{equation}
\mathrm{Var}[\widetilde{\beta}] = \frac{1}{32\eta (2-\eta) n N ^2},
\label{eq:Heisenberg}
\end{equation}
with only a classical a priori knowledge of $\varphi_-$ around a peak of maximal probability.

We remark that in the presence of inefficient detectors, corresponding to $\eta<1$, the variance of the estimator~\eqref{eq:Heisenberg} changes only by a constant factor $1/(\eta (2-\eta))$
with respect to the ideal case ($\eta=1$), meaning that the effect of such losses does not affect the Heisenberg scaling and is easily mitigated by increasing the mean photon number by the square root of the same constant factor (see again Fig.~\ref{fig:prob}). 

It is possible to reverse the roles of the parameters, i.e. once $\beta$ is known with classical precision, we can use the same procedure to estimate $\varphi_-$. To this end, given the measurements $x_1,\dots,x_n$, the maximum likelihood estimator $\widetilde{\varphi_-}$ is now defined by
\begin{equation}
P(\beta,\widetilde{\varphi_-}) = \frac{1}{n}\sum_{i=1}^nx_i.
\label{eq:MLE_def_f}
\end{equation}
Similarly to Eq.~\eqref{eq:sensitivity}, for large $n$ the variance of $\widetilde{\varphi_-}$ can be approximated by the error propagation formula
\begin{align}
\mathrm{Var}[\widetilde{\varphi_-}] \simeq \frac{P(\beta,\varphi_-)\big(1-P(\beta,\varphi_-)\big)}{n(\partial P/\widetilde{\partial\varphi_-})^2}.
\label{eq:sensi_other}
\end{align}  

A plot of $1/ N \mathrm{Var}[\widetilde{\varphi_-}]$ in Fig.~\ref{fig:rescaled_vars} reveals that much like the case of $\beta$, there is a neighbourood of the origin where the sensitivity is maximal, although the scaling with $N$ is classical. Indeed, setting $\beta\simeq k\pi$, $k$ integer, with classical precision and assuming $\varphi_-$ to be in a neighbourhood of $h\pi$ of size $1/ N$, we have
\begin{widetext}
\begin{align}
\mathrm{Var}[\widetilde{\varphi_-}] &\simeq \frac{P(\beta,\varphi_-)\big(1-P(\beta, \varphi_-)\big)}{n(\partial P/{\partial \varphi_-})^2}\bigg|_{\beta=k\pi+\delta\beta,\,\varphi_-=h\pi+\delta\varphi}
 \nonumber\\
&= \frac{1}{4\tilde{\eta}n N }  +\frac{6\tilde{\eta}\Big( \tilde{\eta}(2\delta\varphi^2 N ^2+\delta\beta^2 N )  + (5-\tilde{\eta})\delta\beta^2  \Big)}{24\tilde{\eta}^2  n N ^2} + O\bigg(\frac{1}{ N ^3}\bigg),
\end{align}
\end{widetext}
with $\delta\beta=O(1/\sqrt{N})$ and $\delta\varphi=O(1/ N)$. The second term is $O(1/N^2)$, whence the dominant term is  the first one, and does not depend on $\delta\beta$ or $\delta\varphi$. Therefore our protocol enables the estimation of $\varphi_-$ as well, with a sensitivity scaling at the standard quantum limit.

A final remark is in order. There are other situations where we are able to reach the Heisenberg limit in the estimation of $\beta$. If $\varphi_-$ is not a multiple of $\pi$, but rather $\varphi_-=(k+\frac{1}{2})\pi$ with integer $k$, all the photons end up in the second channel rather than the first. This suggests that in this case it might be possible to obtain the same results as above simply by performing the anti-squeezing operation on the second channel instead of the first one. Indeed, in this case the  detection probability becomes
\begin{widetext}
\begin{align}
P'(\beta,\varphi_-) = \bigg\{1 + \tilde{\eta}\bigg[2 N  + \Big(2\sin(\varphi_-)^2+\tilde{\eta} \cos(\varphi_-)^4\Big) N ^2
- 2\sin(\varphi_-)^2\cos(2\beta) N (1+ N )\bigg]\bigg\}^{-1/2}.
\label{eq:boh_3}
\end{align}
\end{widetext}
We can clearly see that $P'(\beta,\varphi_--\pi/2)=P(\beta,\varphi_-)$, hence as expected we can repeat the procedures outlined before, both for the estimation of $\beta$ and $\varphi_-$, with the only difference that the peaks of $P$ in $(\beta,\varphi_-)=(h,k)\pi$ are now replaced by $(h,k+\frac{1}{2})\pi$, but otherwise identical results.

\section{Discussion}

We have provided a quantum metrological protocol which makes use of an input squeezed vacuum followed by an anti-squeezing operation, as well as quantum interference, as metrological resources to achieve the Heisenberg limit in the estimation of either the average phase in a Mach-Zehnder interferometer, or the relative phase of the two given squeezers. The protocol also allows to estimate the relative phase  between the two arms of the Mach-Zehnder network at the standard quantum limit.

Our protocol shows not only that the `global phase' in a Mach-Zehnder interferometer has a physical effect, which allows it to be detected, but also that in general its knowledge is necessary even if one wishes to estimate other parameters. Indeed, while we choose to focus primarily on the estimation of $\beta$, the local indistinguishability of $(\beta,\varphi_-)$ entails that any attempt at estimating $\varphi_-$ cannot be carried out without some knowledge on the value $\beta$, independently of one's expectations for the sensitivity of the estimation procedure. This might come in the form of a  knowledge of the parameter, as we have done, or for instance in the form of a reasonably justified prior distribution---which would transform the protocol into a Bayesian estimation problem~\cite{stoica01,li12}. Remarkably, despite the local indistinguishability of the parameters, it is possible to determine $\beta$ with Heisenberg precision with only a classical knowledge of $\varphi_-$.

The physical effect of the parameters $\beta$ of $\varphi_-$ is most easily understood in phase space picture. The parameter $\beta$ represents the relative angle between the ellipse associated with the state after the Mach-Zehnder configuration, and the ellipse associated with the output of the interferometer right before detection. On the other hand, $\varphi_-$ controls the squeezing ratio between the first and the second channel of the interferometer. Both these effects contribute to the interference pattern observed in the detection probability $P$ as a function of $\beta$ and $\varphi_-$. The role of constructive quantum interference as a necessary condition for achieving the Heisenberg limit in distributed quantum metrology is a particularly interesting element which emerges from our analysis and will be further explored in future works.

Remarkably, we have shown that the protocol is robust to  detector inefficiency, which in practice could seriously hinder the estimation sensitivity with respect to the ideal case, but actually only reduce the sensitivity by a constant factor, and can thus be overcome by increasing the mean photon number by the same amount.

\section{Acknowledgments}
This work was partially supported by the Office of Naval Research (ONR) Global (Award No. N62909-18-1-2153). D.G. is supported by the University of Portsmouth. 
P.F. is partially supported by Istituto Nazionale di Fisica Nucleare (INFN) through the project ``QUANTUM,'' and by the Italian National Group of Mathematical Physics of Istituto Nazionale di Alta Matematica (GNFM-INdAM). 

\bibliographystyle{plain}

\begin{thebibliography}{9}
\bibitem{caves81}
C. M. Caves, 
Quantum-mechanical noise in an interferometer, 
\href{https://doi.org/10.1103/PhysRevD.23.1693}{Phys. Rev. D \textbf{23}, 1693 (1981).}

\bibitem{yurke86}
B. Yurke, S. L. McCall, and J. R. Klauder, SU(2) and SU(1,1) interferometers, \href{https://doi.org/10.1103/PhysRevA.33.4033}{Phys. Rev. A \textbf{33}, 4033 (1986).} 

\bibitem{holland93}
M. J. Holland and K. Burnett, Interferometric detection of optical phase shifts at the Heisenberg limit, \href{https://doi.org/10.1103/PhysRevLett.71.1355}{Phys. Rev. Lett. \textbf{71}, 1355 (1993).}

\bibitem{giovannetti04}
V. Giovannetti, S. Lloyd, and L. Maccone, Quantum-Enhanced Measurements: Beating the Standard Quantum Limit, \href{https:/doi.org/10.1126/science.1104149}{Science \textbf{306}, 1330 (2004).}

\bibitem{giovannetti06}
V. Giovannetti, S. Lloyd, and L. Maccone, Quantum Metrology, \href{https://doi.org/10.1103/PhysRevLett.96.010401}{Phys. Rev. Lett. \textbf{96}, 010401 (2006).}

\bibitem{giovannetti11}
V. Giovannetti, S. Lloyd, and L. Maccone, Advances in Quantum Metrology, \href{https://doi.org/10.1038/nphoton.2011.35}{Nature Photonics \textbf{5}, 222 (2011).}

\bibitem{pezze14}
L. Pezz\`e and A. Smerzi, Quantum theory of phase estimation, in \textit{Proceedings of the International School of Physics ``Enrico Fermi"}, 691 (2014).

\bibitem{demkowicz15}
R. Demkowicz-Dobraz\'anski, M. Jarzyna, and J. Ko\l ody\'nski, Quantum Limits in Optical Interferometry, \href{https://doi.org/10.1016/bs.po.2015.02.003}{Progress in Optics \textbf{60}, 345 (2015).}

\bibitem{steinert10} 
S. Steinert, F. Dolde, P. Neumann, A. Aird, B. Naydenov, G. Balasubramanian, F. Jelezko, and J. Wrachtrup, High sensitivity magnetic imaging using an array of spins in diamond, \href{https://aip.scitation.org/doi/10.1063/1.3385689}{Rev. Sci. Instrum. \textbf{81}, 043705 (2010).}

\bibitem{hall12} 
L. T. Hall, G. C. G. Beart, E. A. Thomas, D. A. Simpson, L. P. McGuinness, J. H. Cole, J. H. Manton, R. E. Scholten, F. Jelezko, J. Wrachtrup, S. Petrou, and L. C. L. Hollenberg,  High spatial and temporal resolution wide-field imaging of neuron activity using quantum NV-diamond, \href{https://doi.org/10.1038/srep00401}{Sci. Rep. \textbf{2}, 401 (2012).}

\bibitem{pham11}
L. M. Pham1, D .Le Sage, P. L. Stanwix, T. K. Yeung, D. Glenn, A. Trifonov, P. Cappellaro, P. R. Hemmer, M. D. Lukin, H. Park, A. Yacoby, and R. L. Walsworth, Magnetic field imaging with nitrogen-vacancy ensembles, \href{https://doi.org/10.1088/1367-2630/13/4/045021
}{New J. Phys. \textbf{13}, 045021 (2011).}

\bibitem{seo07} 
M. Seo, A. Adam, J. Kang, J. Lee, S. Jeoung, Q. H. Park, P. Planken, and D. Kim, Fourier-transform terahertz near-field imaging of one-dimensional slit arrays: mapping of electric-field-, magnetic-field-, and Poynting vectors, \href{https://doi.org/10.1364/OE.15.011781}{Optics Express \textbf{15}, 11781 (2007).}

\bibitem{baumgratz16} 
T. Baumgratz, and A. Datta, Quantum Enhanced Estimation of a Multidimensional Field, \href{https://doi.org/10.1103/PhysRevLett.116.030801}{Phys. Rev. Lett. \textbf{116}, 030801 (2016).}

\bibitem{humphreys13}
P. C. Humphreys, M. Barbieri, A. Datta, and I. A. Walmsley, Quantum Enhanced Multiple Phase Estimation, \href{https://doi.org/10.1103/PhysRevLett.111.070403}{Phys. Rev. Lett. \textbf{111}, 070403 (2013).}

\bibitem{liu16} 
J. Liu, X.-M. Lu, Z. Sun, and X.Wang, J. Phys. A \textbf{49}, Quantum multiparameter metrology with generalized entangled coherent state, \href{https://doi.org/10.1088/1751-8113/49/11/115302}{J. Phys. A: Math. Theor. \textbf{49}, 115302 (2016).}

\bibitem{yue14} 
J.-D. Yue, Y.-R. Zhang, and H. Fan, Quantum-enhanced metrology for multiple phase estimation with noise, \href{https://doi.org/10.1038/srep05933}{Sci. Rep. \textbf{4}, 5933 (2014).}

\bibitem{knot16} 
P. A. Knott, T. J. Proctor, A. J. Hayes, J. F. Ralph, P. Kok, and J. A. Dunningham, Local versus global strategies in multiparameter estimation, \href{https://doi.org/10.1103/PhysRevA.94.062312}{Phys. Rev. A \textbf{94}, 062312 (2016).}

\bibitem{gagatsos16} 
C. N. Gagatsos, D. Branford, and A. Datta, Gaussian systems for quantum-enhanced multiple phase estimation, \href{https://doi.org/10.1103/PhysRevA.94.042342}{Phys. Rev. A \textbf{94}, 042342 (2016).}

\bibitem{ciampini15} 
M. A. Ciampini, N. Spagnolo, C. Vitelli, L. Pezz\'e, A. Smerzi, and F. Sciarrino, Quantum-enhanced multiparameter estimation in multiarm interferometers, \href{https://doi.org/10.1038/srep28881}{Sci. Rep. \textbf{6}, 28881 (2016).}

\bibitem{eldredge18}
Z. Eldredge, M. Foss-Feig, J. A. Gross, S. L. Rolston, and A. V. Gorshkov, Optimal and secure measurement protocols for quantum sensor networks, \href{https://doi.org/10.1103/PhysRevA.97.042337}{Phys. Rev. A \textbf{97}, 042337 (2018).}

\bibitem{arai15}
K. Arai, C. Belthangady, H. Zhang, N. Bar-Gill, S. J. De-Vience, P. Cappellaro, A. Yacoby, and R. L. Walsworth, Fourier magnetic imaging with nanoscale resolution and compressed sensing speed-up using electronic spins in diamond,
\href{https://doi.org/10.1038/nnano.2015.171}{Nat. Nanotechnol. \textbf{10}, 859 (2015).}

\bibitem{lazariev15}
A. Lazariev and G. Balasubramanian, A nitrogen-vacancy spin based molecular structure microscope using multiplexed projection reconstruction, \href{https://doi.org/10.1038/srep14130}{Sci. Rep. \textbf{5}, 14130 (2015).}

\bibitem{komar14}
P. Komar, E. M. Kessler, M. Bishof, L. Jiang, A. S. S\o rensen, J. Ye, and M. D. Lukin, A quantum network of clocks, \href{https://doi.org/10.1038/nphys3000}{Nat. Phys. \textbf{10}, 582 (2014).}

\bibitem{proctor17}
T. J. Proctor, P. A. Knott, and J. A. Dunningham, Multiparameter Estimation in Networked Quantum Sensors, \href{https://doi.org/10.1103/PhysRevLett.120.080501}{Phys. Rev. Lett. \textbf{120}, 080501 (2018).}

\bibitem{boixo07}
S. Boixo, S. T. Flammia, C. M. Caves, and JM Geremia, Generalized Limits for Single-Parameter Quantum Estimation, \href{https://doi.org/10.1103/PhysRevLett.98.090401}{Phys. Rev. Lett. \textbf{98}, 090401 (2007).}

\bibitem{lang13}
M. D. Lang and C. M. Caves, Optimal Quantum-Enhanced Interferometry Using a Laser Power Source, \href{https://doi.org/10.1103/PhysRevLett.111.173601}{Phys. Rev. Lett. \textbf{111}, 173601 (2013).}

\bibitem{zhuang18}
Q. Zhuang, Z. Zhang, and J. H. Shapiro, Distributed quantum sensing using continuous-variable multipartite entanglement, \href{https://doi.org/10.1103/PhysRevA.97.032329}{Phys. Rev. A \textbf{97}, 032329 (2018).}

\bibitem{guo19}
X. Guo, C. R. Breum, J. Borregaard, S. Izumi, M. V. Larsen, T. Gehring, M. Christandl, and J. S. Neergaard-Nielsen, U. L. Andersen, Distributed quantum sensing in a continuous variable entangled network, \href{https://doi.org/10.1038/s41567-019-0743-x}{Nat. Phys. \textbf{16}, 281 (2020).}

\bibitem{gramegna20}
G. Gramegna, D. Triggiani, P. Facchi, F. A. Narducci, V. Tamma, Heisenberg scaling precision in multi-mode distributed quantum metrology, preprint arXiv:2003.12550 [quant-ph].

\bibitem{gramegna21}
G. Gramegna, D. Triggiani, P. Facchi, F. A. Narducci, and V. Tamma, Typicality of Heisenberg scaling precision in multimode quantum metrology, \href{https://doi.org/10.1103/PhysRevResearch.3.013152}{Phys. Rev. Research \textbf{3}, 013152 (2021)}.

\bibitem{schleich}
W. P. Schleich, \href{https://doi.org/10.1002/3527602976}{\textit{Quantum Optics in Phase Space}}, Wiley‐VCH Verlag Berlin GmbH (2001).

\bibitem{gatto20}
D. Gatto, P. Facchi, F. A. Narducci, and V. Tamma, Distributed quantum metrology with a single squeezed-vacuum source, \href{https://doi.org/10.1142/S0219749919410193}{Int. J. Quantum Inf. \textbf{18},  1941019 (2020).}

\bibitem{gatto19-2}
D. Gatto, P. Facchi, and V. Tamma, Phase space Heisenberg-limited estimation of the average phase shift in a Mach–Zehnder interferometer, \href{https://doi.org/10.1103/PhysRevResearch.1.032024}{Phys. Rev. Res. \textbf{1}, 032024(R) (2019).}

\bibitem{stoica01}
P. Stoica, and T. L. Marzetta, Parameter estimation problems with singular information matrices, \href{https://doi.org/10.1109/78.890346}{IEEE Trans. Signal Process \textbf{49}, 87 (2001).}

\bibitem{stoica82}
P. Stoica and T. S\"oderstr\"om, On non-singular information matrices and local identifiability, \href{https://doi.org/10.1080/00207178208932896}{Int. J. Contr. \textbf{36}, 323 (1982).}

\bibitem{li12}
Y.-H. Li, P.-C. Yeh, An Interpretation of the Moore-Penrose Generalized Inverse of a Singular Fisher Information Matrix, \href{10.1109/TSP.2012.2208105}{IEEE Trans. Signal Process \textbf{60}, 5532 (2012).}

\bibitem{rothenberg71}
T. J. Rothenberg, Identification in Parametric Models, \href{https://doi.org/10.2307/1913267}{Econometrica \textbf{39}, 577 (1971).}
\end{thebibliography}

\newpage

\onecolumn
\appendix
\section{Squeezing of the state at the two output channels of the Mach-Zehnder network}
\label{sec:append_00}
The lengths of the semi-axes of the ellipses describing the marginal Wigner functions in Eq.~\eqref{eq:reduced_wigners} at the two channels 1 and 2 of the Mach-Zehnder network as shown in Figs.~\ref{fig:diffphase} and~\ref{fig:sumphase} can be calculated by expressing the covariance matrix $\sigma=\sigma_\mathrm{MZ}$ in Eq.~\eqref{eq:interferometer_cov_transf} in block form as
\begin{equation}
\sigma_\mathrm{MZ} = \begin{pmatrix}
\sigma_1 & \tau \\
\tau^T & \sigma_2
\end{pmatrix},
\end{equation}
where
\begin{align}
\sigma_1 &= \begin{pmatrix}
s^{(1)}_+ & \cos(\varphi_+)\sin(2\varphi_++2\theta_\mathrm{in})^2\sqrt{N(1+N)} \\
\cos(\varphi_+)\sin(2\varphi_++2\theta_\mathrm{in})^2\sqrt{N(1+N)} & s^{(1)}_-
\end{pmatrix} , \nonumber\\
\sigma_2 &= \begin{pmatrix}
s^{(2)}_- & -\sin(\varphi_-)^2\sin(2\varphi_++2\theta_\mathrm{in})\sqrt{N(1+N)} \\
\sin(\varphi_-)^2\sin(2\varphi_++2\theta_\mathrm{in})\sqrt{N(1+N)} & 
s^{(2)}_+
\end{pmatrix},  \nonumber\\
\tau &= \frac{1}{2}\begin{pmatrix}
-\sin(2\varphi_-)\sin(2\varphi_++2\theta_\mathrm{in})\sqrt{N(1+N)} & \tau_-, \nonumber\\
\tau_+ \sin(2\varphi_-)\sin(2\varphi_++2\theta_\mathrm{in})\sqrt{N(1+N)}
\end{pmatrix},
\end{align} 
and
\begin{align}
&s^{(1)}_\pm = \frac{1}{2}(1+2\cos(\varphi_-)^2\sqrt{N}(\sqrt{N}\pm\cos(2\varphi_++2\theta_\mathrm{in})\sqrt{1+N})), \nonumber\\
&s^{(2)}_\pm = \frac{1}{2}(1+2\sin(\varphi_-)^2\sqrt{N}(\sqrt{N}\pm 2\cos(2\varphi_++\theta_\mathrm{in})\sqrt{1+N})), \nonumber\\
&\tau_\pm = \sin(2\varphi_-)\sqrt{N}(\cos(2\varphi_++2\theta_\mathrm{in})\sqrt{1+N}\pm\sqrt{N}).
\end{align}

The lengths of the two semi-axes of the marginal Wigner function $W_1(x_1, p_1; \sigma_\mathrm{MZ})$ at the first channel are given by the eigenvalues of $\sigma_1$ and reads
\begin{align}
1+\cos(\varphi_-)^2\Bigl( N \pm\sqrt{ N (1+ N )}\Bigr),
\end{align}
as reported in Eq.~\eqref{eq:prop_1} in the text, whereas the lengths of the two semi-axes of the marginal Wigner function $W_2(x_2, p_2; \sigma_\mathrm{MZ})$  at the second channel are given by the eigenvalues of $\sigma_2$
\begin{align}
1+\sin(\varphi_-)^2\Bigl( N \pm\sqrt{ N (1+ N )}\Bigr),
\end{align}
which in turn gives Eq.~\eqref{eq:prop_2}.

\section{Linear optical networks in phase space}
\label{sec:append_0}
A passive linear optical network is described by a unitary operator $\hat{U}$ on the underlying Hilbert space of the $2$-mode electromagnetic field, which acts on the annihilation operators according to the relation
\begin{equation}
\hat{U}^\dag\hat{a}_i\hat{U} = \sum_{j=1}^2{\cal U}_{ij}\,\hat{a}_j,
\end{equation}
where $\mathcal{U}$ is a $2\times 2$ unitary matrix.

Consider the vector of the quadratures $\hat{\mathbf{R}}=(\hat{x}_1,\hat{p}_1,\hat{x}_2,\hat{p}_2)^T$, with $\hat{x}_j = (\hat{a}_j+\hat{a}^\dag_j)/\sqrt{2}$ and  $\hat{p}_j = (\hat{a}_j-\hat{a}^\dag_j)/\sqrt{2} i$. 
By defining the unitary $4\times 4$ matrix
\begin{equation}
W = {1\over\sqrt{2}}\bigoplus_{j=1}^2\begin{pmatrix}
1 & i \\
1 & -i
\end{pmatrix}, 
\end{equation}
we have the identity
\begin{equation}
\hat{\mathbf{A}} := (\hat{a}_1,\hat{a}^\dag_1,\hat{a}_2,\hat{a}^\dag_2)^T = W\hat{{\bf R}}.
\end{equation}
Thus, the network transforms the quadratures according to
\begin{align}
 \hat{U}^\dag\hat{R}_i\hat{U} &=  
 \hat{U}^\dag\sum_{j=1}^{2}(W^\dag_{i,2j-1}\hat{a}_j + W^\dag_{i,2j}\hat{a}^\dag_j)\hat{U} = \sum_{j,k=1}^2(W^\dag_{i,2j-1}{\cal U}_{jk}\hat{a}_k + W^\dag_{i,2j}{\cal U}^*_{jk}\hat{a}^\dag_k)  \nonumber\\
 &=\sum_{h=1}^{4}\sum_{j,k=1}^2(W^\dag_{i,2j-1}{\cal U}_{jk}W_{2k-1,h} + W^\dag_{i,2j}{\cal U}^*_{jk}W_{2k,h})\hat{R}_h := \sum_{h=1}^{4}O_{ih}\hat{R}_h,
\end{align}
that is
\begin{equation}
 \hat{U}^\dag\hat{\mathbf{R}} \hat{U} = O \hat{\mathbf{R}}.
\end{equation}
The matrix $O$ has the form
\begin{equation}
 O= W^\dag\begin{pmatrix}
           {\cal U}_{11} & 0 &  \mathcal{U}_{12} & 0 \\
           0 & {\cal U}^*_{11} & 0 & \mathcal{U}^*_{12} \\
           \mathcal{U}_{21} & 0 &  {\cal U}_{22} & 0 \\
           0 & \mathcal{U}^*_{21} &  0 & {\cal U}^*_{22} \\
          \end{pmatrix}W,
\end{equation}
which, as can  be easily verified, is both orthogonal and symplectic, and can be simplified  as follows
\begin{equation}
  O= \begin{pmatrix}
           \Re({\cal U}_{11})\openone_2-i\Im({\cal U}_{11})\sigma_y  & \Re({\cal U}_{12})\openone_2-i\Im({\cal U}_{12})\sigma_y \\
           \\
           \Re({\cal U}_{21})\openone_2-i\Im({\cal U}_{21})\sigma_y & \Re({\cal U}_{22})\openone_2-i\Im({\cal U}_{22})\sigma_y
          \end{pmatrix}.
          \label{eq:notaz_paolo}
\end{equation}

The 2-mode covariance matrix 
is defined as $\sigma= (\sigma_{j k})$ where
\begin{equation}
\sigma_{j k} = \frac{1}{2} \bigl\langle \{\hat{R}_j , \hat{R}_k\}  \bigr\rangle - \bigl\langle  \hat{R}_j   \bigr\rangle \, \bigl\langle  \hat{R}_k   \bigr\rangle,
\end{equation}
with the expectation value taken on the field state.
Therefore, if the field has initially covariance $\sigma_{\mathrm{in}}$, after the action of a passive linear optical network it has a covariance
given by
\begin{equation}
\sigma_{\mathrm{out}} =  O\sigma_{\mathrm{in}}\,O^T,
\end{equation}
which is nothing but a (symplectic) rotation in phase space.

 In the case of the Mach-Zehnder interferometer, the unitary matrix $\mathcal{U}$ matrix reads
\begin{equation}
\mathcal{U}_{\mathrm{MZ}} = \frac{1}{2}\begin{pmatrix}
1 & -i \\
-i & 1
\end{pmatrix} \begin{pmatrix}
e^{i\varphi_1} & 0 \\
0 & e^{i\varphi_2}
\end{pmatrix} \begin{pmatrix}
1 & i \\
i & 1
\end{pmatrix} = e^{i\varphi_+}\begin{pmatrix}
\cos(\varphi_-) & -\sin(\varphi_-) \\
\sin(\varphi_-) & \cos(\varphi_-)
\end{pmatrix},
\end{equation}
hence the corresponding rotation in phase space is given by
\begin{equation}
O_\mathrm{MZ} = \begin{pmatrix}
\cos(\varphi_+)\cos(\varphi_-) & -\sin(\varphi_+)\cos(\varphi_-) & -\cos(\varphi_+)\sin(\varphi_-) & \sin(\varphi_+)\sin(\varphi_-) \\
\sin(\varphi_+)\cos(\varphi_-) & \cos(\varphi_+)\cos(\varphi_-) & -\sin(\varphi_+)\sin(\varphi_-) & -\cos(\varphi_+)\sin(\varphi_-) \\
\cos(\varphi_+)\sin(\varphi_-) & -\sin(\varphi_+)\sin(\varphi_-) & \cos(\varphi_+)\cos(\varphi_-) & -\sin(\varphi_+)\cos(\varphi_-)\\
\sin(\varphi_+)\sin(\varphi_-) & \cos(\varphi_+)\sin(\varphi_-) & \sin(\varphi_+)\cos(\varphi_-) & \cos(\varphi_+)\cos(\varphi_-)
\end{pmatrix},
\end{equation}
which can be written more compactly as in Eq.~\eqref{eq:OMZ} in the text.

\section{Covariance matrices in presence of losses}
\label{sec:append_1}
A photon loss associated with a given mode $S$, described by the density operator $\hat{\rho}$ and the corresponding covariance matrix $\sigma$, can be modelled by making use of a fictitious beam splitter with transmittivity $\eta$. The external environment on the second channel of the beam splitter is assumed to be in the vacuum state, while the output state in the first, `physical' channel, is obtained by tracing on the second one. The quantum channel $\Phi_\eta$ obtained in this way is called the attenuator channel,
\begin{equation}
\Phi_\eta(\hat{\rho}) = \mathrm{Tr}(\hat{U}_\mathrm{BS}\,\hat{\rho}\otimes\ket{0}_E\bra{0}\hat{U}^\dag_\mathrm{BS})_E,
\label{eq:boh_buh}
\end{equation}
where the subscript $E$ denotes the environment and $\hat{U}_\mathrm{BS}$ is the unitary operator describing a beam splitter. As shown in the previous section, the beam splitter is described by the unitary matrix
\begin{equation}
\mathcal{U}_{\mathrm{BS}}(\alpha) = \begin{pmatrix}
\cos(\alpha) & -i\sin(\alpha) \\
-i\sin(\alpha) & \cos(\alpha)
\end{pmatrix} = e^{i \alpha \sigma_x},
\end{equation}
where $\eta=\cos(\alpha)^2$, corresponding---by using Eq.~\eqref{eq:notaz_paolo}---to the symplectic rotation  in phase space	
\begin{equation}
O_{\mathrm{BS}}(\alpha) = \begin{pmatrix}
 \cos(\alpha)\, \openone_2 & i \sin(\alpha)\, \sigma_y \\
i \sin(\alpha)\, \sigma_y  & \cos(\alpha)\, \openone_2 \end{pmatrix}.
\end{equation}
Thus, the covariance matrix
\begin{equation}
\sigma_\mathrm{SE} = \begin{pmatrix}
\sigma & 0 \\
 0 & \frac{1}{2} \openone_2
\end{pmatrix} ,
 \end{equation}
describing  the system $S$, with covariance matrix $\sigma$, and the environment $E$ in the vacuum state, is transformed by the interaction with the beam splitter as
 \begin{equation}
O_{\mathrm{BS}} (\alpha)\, \sigma_\mathrm{SE}\,  O_{\mathrm{BS}}(\alpha)^T  = 
  \begin{pmatrix}
\cos(\alpha)^2\, \sigma +\frac{1}{2} \sin(\alpha)^2\, \openone_2  & i \sin(\alpha)\cos(\alpha) \left( \sigma - \frac{1}{2}\openone_2 \right) \sigma_y \\
 -i \sin(\alpha) \cos(\alpha)\, \sigma_y \left( \sigma - \frac{1}{2}\openone_2\right)    & \frac{1}{2} \cos(\alpha)^2\, \openone_2 + \cos(\alpha)^2\, \sigma_y \sigma \sigma_y.
\end{pmatrix}.
 \end{equation}
Therefore a photon loss, described by the quantum channel in Eq.~\eqref{eq:boh_buh}, simply maps the (physical) covariance matrix into the covariance matrix of  a convex combination of the signal with vacuum fluctuations,
\begin{equation}
\sigma \longmapsto  \cos(\alpha)^2\, \sigma +\frac{1}{2} \sin(\alpha)^2\, \openone_2 = \eta\,\sigma + \frac{1-\eta}{2}\openone_2.
\end{equation}
The transformation above easily generalises to an $M$-mode Gaussian state going through $M$ lossy lines with equal loss, $\Phi^{\otimes M}_\eta$, as
\begin{equation}
\sigma \longmapsto \eta\,\sigma + \frac{1-\eta}{2}\openone_{2M}.
\label{eq:boh}
\end{equation}

\section{Detection probabilities in the Mach-Zehnder interferometer}
\label{sec:append_2}
The detection probability of our measurement, in the presence of ideal detectors, is given by 
\begin{align}
P &= \mathrm{Tr}(\hat{S}_1(z_\mathrm{out})|00\rangle\langle00|\hat{S}_1^\dag(z_\mathrm{out})\hat{\rho}_\mathrm{MZ}) = (2\pi)^2 \int W_{\sigma_\mathrm{out}}(\boldsymbol\xi) \, W_{\sigma_\mathrm{MZ}}(\boldsymbol\xi) \,\diff^4\xi
\label{eq:49}
\end{align}
with $\sigma_\mathrm{out}$ given in Eq.~\eqref{eq:cov_meas}, 
\begin{equation}
\sigma_\mathrm{MZ}=O_\mathrm{MZ}\sigma_\mathrm{in}O_\mathrm{MZ}^T,
\end{equation} 
and $\sigma_\mathrm{in}$ given in Eq.~\eqref{eq:squeezing_in}.  Thus $P$ can also be expressed as 
\begin{align}
P &= \mathrm{Tr}(|00\rangle\langle00|\hat{S}_1^\dag(z_\mathrm{out})\hat{\rho}_\mathrm{MZ}\hat{S}_1(z_\mathrm{out})) = (2\pi)^2 \int W_{\sigma_\mathrm{vac}}(\boldsymbol\xi) \, W_{S_1(z_\mathrm{out})^{-1}\sigma_\mathrm{MZ}S_1(z_\mathrm{out})^{-1}}(\boldsymbol\xi) \,\diff^4\xi ,
\label{eq:50}
\end{align}
where $\sigma_\mathrm{vac}=\frac{1}{2}\openone_4$ is the covariance matrix of the 2-mode vacuum state and 
\begin{equation}
S_1(z) = 
\begin{pmatrix}
S(z) & 0 \\
 0 & \openone_2
\end{pmatrix} ,
\end{equation}
with $S(z)$ defined in Eq.~\eqref{eq:squeezer},
 \begin{equation}
 S(z) = S(r  e^{i\theta}) = e^{i \theta \sigma_y} 
 \begin{pmatrix}
e^{r} & 0 \\
 0 & e^{-r}
\end{pmatrix}  e^{-i \theta \sigma_y}.
 \end{equation}
A Gaussian integration of~\eqref{eq:50} gives
\begin{equation}
P = \frac{1}{\sqrt{\det \left(\frac{1}{2}\openone_4+ S_1(z_\mathrm{out})^{-1} \sigma_\mathrm{MZ}S_1(z_\mathrm{out})^{-1}\right)}}.
\label{eq:boh_2}
\end{equation}

If we consider two  detectors with quantum efficiency $\eta$, modelled by attenuator channels right before them,  the probability~\eqref{eq:49} becomes
\begin{align}
P = \mathrm{Tr}\Big(\ket{00}\bra{00}\Phi_\eta\otimes\Phi_\eta\big(\hat{S}_1^\dag(z_\mathrm{out})\hat{\rho}_\mathrm{MZ}\hat{S}_1(z_\mathrm{out})\big)\Big),
\end{align}
where $\Phi_\eta$ is the attenuator channel defined in Eq.~\eqref{eq:boh_buh}. Thus applying the transformation~\eqref{eq:boh} to the second term in the determinant of Eq.~\eqref{eq:boh_2}, we obtain
\begin{align}
P &= \frac{1}{\sqrt{\det\left(\frac{1}{2}\openone_4+ \eta S_1(z_\mathrm{out})^{-1}  \sigma_\mathrm{MZ}S_1(z_\mathrm{out})^{-1} + \frac{1-\eta}{2}\openone_4\right)}} = \nonumber\\
&= \det\left( \frac{2-\eta}{2} \openone_4+ \eta S_1(z_\mathrm{out})^{-1} \sigma_\mathrm{MZ}S_1(z_\mathrm{out})^{-1} \right)^{-\frac{1}{2}} \nonumber\\
&= \det(S_1(z_\mathrm{out}))^{-\frac{1}{2}} \det\left( \frac{2-\eta}{2} S_1(z_\mathrm{out})^2 + \eta \sigma_\mathrm{MZ}\right)^{-\frac{1}{2}}   \det(S_1(z_\mathrm{out}))^{-\frac{1}{2}} \nonumber\\
&= \det\big( (2-\eta)\,\sigma_\mathrm{out} + \eta\,\sigma_\mathrm{MZ} \big)^{-1/2},
\end{align}
where in the last equality we used $\det(S_1(z_\mathrm{out})) = \det(S(z_\mathrm{out})) = 1$, and $\frac{1}{2} S_1(z_\mathrm{out})^2= \sigma_\mathrm{out}$.

By making use of expressions~\eqref{eq:interferometer_cov_transf},~\eqref{eq:OMZ} and~\eqref{eq:squeezing_in} for $\sigma_\mathrm{MZ}$, $O_\mathrm{MZ}$ and $\sigma_\mathrm{in}$, after some tedious but simple algebra 
we finally get
 \begin{align}
P(\beta,\varphi_-) = \bigg\{1 + \tilde{\eta}\bigg[2 N  + \Big(2\cos(\varphi_-)^2 + \tilde{\eta}\sin(\varphi_-)^4\Big) N ^2
- 2&\cos(\varphi_-)^2\cos(2\beta)\, N (1+ N )\bigg]\bigg\}^{-1/2}, 
\label{eq:main_prob}
\end{align}
as claimed in the main text. 

The counterpart of the previous probability when the anti-squeezing is performed on the second channel can be calculated with the analogous equation
\begin{align}
P'(\beta,\varphi_-) &= \mathrm{Tr}\Big(\ket{00}\bra{00}\Phi_\eta\otimes\Phi_\eta\big(\hat{S}_2^\dag(z_\mathrm{out})\hat{\rho}_\mathrm{MZ}\hat{S}_2(z_\mathrm{out})\big)\Big) = \nonumber\\
&= \bigg\{1 + \tilde{\eta}\bigg[2 N  + \Big(2\sin(\varphi_-)^2+\tilde{\eta}\cos(\varphi_-)^4\Big) N ^2
- 2\sin(\varphi_-)^2\cos(2\beta) N (1+N)\bigg]\bigg\}^{-1/2},
\label{eq:prob_lossy_detec_mode2}	
\end{align} 
where $\hat{S}_2(z) = e^{\frac{1}{2}(z\hat{a}_2^{\dag2}-z^*\hat{a}_2^2)}$. This gives Eq.~\eqref{eq:boh_3} in the main text.

\section{Localisation of the probability peaks}
\label{sec:append_3}
The diameters of a given level curve of the probability $P$ in Eq.\eqref{eq:main_prob} are found simply by intersecting it with each of the coordinate axes. Let $P_0\in(0,1)$ be fixed, then the intersections with the $\beta$-axis are found from the equation
 \begin{align}
P(\beta_*,0) = \Big[1 + 4\tilde{\eta}N (1+ N )\sin(\beta_*)^2\Big]^{-1/2} = P_0, 
\end{align}
which immediately yields
\begin{equation}
\sin(\beta_*)^2 = \frac{1-P_0^2}{4\tilde{\eta} N (1+ N )P_0^2},
\end{equation}
as claimed in Eq.~\eqref{eq:boh_4} in the text.
Analogously, the intersections with the $\varphi_-$-axis are the solutions to
\begin{align}
P(0,\varphi_*) = \Big[1 + \tilde{\eta}^2N^2\sin(\varphi_*)^4 + 2\tilde{\eta}N\sin(\varphi_*)^2 \Big]^{-1/2} = P_0. 
\end{align}
This is a second-degree equation for $\sin(\varphi_*)^2$,
\begin{align}
\tilde{\eta}^2N^2\sin(\varphi_*)^4 + 2\tilde{\eta}N\sin(\varphi_*)^2 -\frac{1-P_0^2}{P_0^2} = 0, 
\end{align}
which admits the solution
\begin{equation}
\sin(\varphi_*)^2 = \frac{\sqrt{P_0^2+\tilde{\eta}(1-P_0^2)} -P_0}{\tilde{\eta} N P_0},
\end{equation}
i.e. Eq.~\eqref{eq:boh_5} in the main text.
\end{document}